\RequirePackage{fix-cm}
\documentclass[smallextended]{svjour3}       
\smartqed  
\usepackage{graphicx}
\usepackage{gensymb}
\usepackage{hyperref}

\usepackage{amsmath}
\usepackage{amsfonts}

\usepackage[round]{natbib}

\usepackage{subcaption}
\usepackage{amssymb}
\usepackage{bm}

\usepackage{multicol}
\usepackage{multirow}

\usepackage{natbib}

\hypersetup{
    colorlinks,
    urlcolor=black,
    citecolor=blue
}

\newcommand\bigcdot[1][.5]{\mathbin{\vcenter{\hbox{\scalebox{#1}{$\bullet$}}}}}

\def\ie{{{i.e.}, }}
\def\eg{{{e.g.}, }}

\def\JON#1{{#1}}

\begin{document}

\title{Gradient boosting with extreme-value theory for wildfire prediction 
}

\titlerunning{Gradient boosting with extreme-value theory}        

\author{Jonathan Koh 
}


\institute{ Institute of Mathematics, EPFL \\ \at
Institute of Mathematical Statistics and Actuarial Science, Oeschger Centre for Climate Change Research, University of Bern \at
               \\
             \email{jonathan.koh@stat.unibe.ch}
}

\date{First version: 18/10/2021/ This version: 31/10/2022}

\maketitle
\begin{abstract}
This paper details the approach of the team \textit{Kohrrelation} in the 2021 Extreme Value Analysis data challenge, dealing with the prediction of wildfire counts and sizes over the contiguous US. Our approach uses ideas from extreme-value theory in a machine learning context with theoretically justified loss functions for gradient boosting. 
We devise a spatial cross-validation scheme and show that in our setting it provides a better proxy for test set performance than naive cross-validation. The predictions are benchmarked against boosting approaches with different loss functions, and perform competitively in terms of the score criterion, finally placing second in the competition ranking. \\

\keywords{Cross-validation  \and Generalized Pareto distribution \and Gradient boosting \and Loss likelihood \and Machine learning \and Wildfire prediction}
\subclass{62G32 \and 62J99 \and 62P12}
\end{abstract}

\section{Introduction}
Wildfires occur in every season of the year and are a natural phenomenon of the forest ecosystem, important for clearing out decayed vegetation and helping plants to reproduce. However, they have the potential to become conflagrations---intense, destructive fires---that may have huge environmental and ecological impacts. Apart from human casualties, these fires can lead to substantial economic losses; global insured claims due to wildfire events have increased dramatically in recent years, from below $\$10$ billion in 2000--2009 to $\$45$ billion in the subsequent decade\footnote{https://www.swissre.com/risk-knowledge/mitigating-climate-risk/yet-more-wildfires.html}.



Wildfires are complex dynamic processes: their occurrences and behaviour are the product of interconnected factors that include the ignition source, fuel composition, topography and the weather. For example, the wind plays a big role in the spread and ease of fire containment, but its effect is magnified in the presence of accumulated biomass on a hilly boreal forest after a prolonged dry spell. The modelling of wildfires is made even more complicated by the need to model the Wildland-to-Urban interface \citep{Stewart.2007}, as $90\%$ of fires are caused by human activity.  

An important measure of wildfire impact and size is the burned area of wildfire events, commonly used by government agencies and aggregated at different spatial and temporal scales for reporting purposes, e.g., \citet{NIFC}. \JON{Although there is a positive relationship between wildfire counts and sizes, with perfect agreement of the lowest value of zero in both variables, more wildfire occurrences in a region do not necessarily translate to higher burned areas.} 
In 2020, nearly 26,000 wildfires burned approximately 9.5 million acres (ac) in the west, compared with the over 33,000 fires that burned just under 0.7 million ac in the east. 
Similarly, although the numbers of wildfires have fallen since the 1990s, the average annual acreage burned since 2000 has more than doubled.


Many statistical approaches have been developed to aid in wildfire prevention and risk mitigation, with most studies modelling wildfire occurrences and sizes separately \citep[][]{taylor.2013, Xi2019,Pereira2019, Jain.2020}, though models that identify latent factors affecting both have been proposed \citep[e.g.,][]{Koh.2021}. Point processes are natural models for the spatiotemporal pattern of occurrences \citep[][]{Peng2005,Genton2006,Tonini2017,Opitz2020}. \citet{Cumming2001}, \citet{Cui2008} and \citet{Pereira2019} suggested modelling fire sizes with various probability distributions. As data usually show heavy-tailed behaviour, only a small fraction of wildfires account for the vast majority of the area burned. Obvious candidates to capture this stem from extreme-value theory, such as the generalized Pareto distribution (GPD) for modelling threshold exceedances \citep[][]{DeZeaBermudez2009,Turkman2010,Pereira2019}. 

Both Bayesian and frequentist methods have been used for explanatory modeling, the former predominantly for hierarchical mixed effect models \citep{Joseph2019,Pimont2020,Koh.2021} and the latter within the generalized additive modelling \citep[GAM,][]{wood2006} framework \citep[][]{Preisler.2004, Brillinger.2006, Vilar.2010, Woolford.2011}. 
Covariates include weather variables such as humidity, temperature, precipitation and meteorologically-based fire danger indices such as the Canadian Fire Weather Index \citep{vanWagner1977}. When available, land-use or locally observed anthropogenic variables like population density and the distance to the nearest train line are used to help to explain human-induced occurrences; spatiotemporal random effects have been incorporated as surrogates for these variables.


If accurate prediction is of primary interest, then machine learning (ML) techniques offer an attractive alternative to the statistical modelling approaches described above. Since the 1990s, the surge in the availability of data and covariates has spurred the use of these techniques to predict wildfire behaviour.
\citet[][]{Jain.2020} found 127 journal papers or conference proceedings published up to the end of 2019 on ML applied to fire occurrence, susceptibility and risk; of these adversarial neural networks (ANN) were the most prominent \citep[\eg][]{Dutta.2013, Shidik.2014,Liang.2019}. For wildfire occurrences, most studies focus on classification tasks instead of count modelling. \citet{Sakr.2010} used metereological variables with support vector machines to predict a four-class fire risk index based on the daily number of fires in Lebanon. \citet{Dutta.2013} compared ten ANN based cognitive imaging systems to determine the relationship between monthly fire incidence and climate for Australia. \citet{Mitsopoulos2017} and \citet{Xie.2018} showed that ensemble learning methods like random forests and boosting trees performed well in estimating area burned or classifying wildfire sizes in Greece and Portugal, respectively. 




Gradient boosting techniques \citep{friedman.2001} have exploded in popularity over the last decade, in part due to the development and dissemination of open-source packages such as \texttt{gbm} \citep{greenwell.gbm} and \texttt{xgboost} \citep{Chen.2016}. 
A key ingredient of gradient boosting is the loss function used to train these models, and choices for these functions have largely been restricted to those that emphasize good prediction of the distributional bulk instead of the tails. For example, squared loss, the default when modelling wildfire sizes, implicitly presupposes normality of the response given the covariates, which may be inappropriate if the focus is predominantly on extreme values. The Poisson loss is popular for modelling wildfire counts within a grid cell, but the zero-inflated nature and potentially heavy tails of count distributions suggest that this loss may be unsuitable. Evaluation metrics should also reflect the non-linear impact of wildfire events; e.g., in many cases, predicting a false negative occurrence is much costlier than predicting a false positive.


As ML methods are prone to overfitting, it is imperative to evaluate models with held-out datasets using robust validation schemes. 
A realistic approach in the forecasting context (when there are no trends) is to leave out the most recent portion of the dataset \citep[\eg][]{Woolford.2011, Dutta.2013, Joseph2019, Koh.2021}. \citet{Dutta.2013} explored different combinations of training-testing splits to identify the best possible paradigm to maximize the generalization capability of their ANN architecture. $K$-fold cross-validation is also popular \citep[][]{Shidik.2014, DeAngelis.2015, Mitsopoulos2017, Xie.2018}, but it may give overly optimistic evaluations for spatially dependent data \citep{Roberts.2017}. An alternative 
is spatial cross-validation \citep{Pohjankukka.2017}, but it is still unclear how best to construct spatial folds in this context, and doing so anyway ignores relevant inter-variable or time dependencies. 





Our work aims to tackle the limitations of the studies mentioned above, and does so in the context of the Extreme Value Analysis 2021 data challenge \citep{data.challenge}. We develop novel gradient boosting models trained with loss functions appropriate for predicting extreme values. Our chosen model for fire counts is a discrete generalized Pareto distribution \citep[dGPD, ][]{shimura.2012} relying on a covariate-dependent parameter that models a chosen high quantile of the distribution, and a shape hyperparameter selected by cross-validation. \JON{The dGPD provides added flexibility in modelling the upper tail of the count distribution, especially when compared to other more standard count distributions like the Poisson.} The model for fire sizes has three components and covariate-dependent probabilities. The first component models the probability of observing no fires, and the others model the probabilities of observing medium-sized and extreme fires. We approximate the conditional distribution above a high threshold with a GPD, and the conditional distribution below the threshold with a truncated log-gamma distribution. 

To improve our models, we also engineered new covariates that incorporate more spatial information into the climatic and land-use covariates provided by averaging them across neighbouring grid cells each month. With a smart imputation method for replacing missing data, we also use the wildfire counts as a covariate when predicting wildfire sizes, and vice versa.



We develop a spatiotemporal cross-validation scheme that provides a better proxy for our models' test set performance than the naive scheme. This involves fitting a space-time latent Gaussian model to pseudo-binary observations that indicate whether a grid cell was masked in a particular month, and then simulating from the fitted model to generate folds of train-test regimes. 





In the remainder of the paper, we first explore the data on wildfires and their covariates, and then introduce the problem set out by the data challenge in \S\ref{sec:data_boosting}.
 We provide general background on extreme-value theory and gradient boosting and on how to combine them in \S\ref{sec:methods-general_boosting}. Our spatiotemporal cross-validation scheme is developed in \S\ref{sec:methods:cv} and
the specific model structure is detailed in \S\ref{sec:model_all}. We highlight the prediction of wildfire activity components in \S\ref{sec:results_boosting}, and compare them to related and competing approaches. We conclude with a discussion and outlook 
in \S\ref{sec:conclusion_boosting}.

\section{Data and exploratory analyses}\label{sec:data_boosting}


The Extreme Value Analysis 2021 data challenge dealt with the prediction of monthly wildfire counts and burned areas at 3503 grid cells across the contiguous US over the period 1993--2015. As fuel moisture is an integral of past precipitation and evaporation mediated by soil field capacity, temporal scales longer than hourly or daily (e.g., monthly in our case) are appropriate for predicting fire risk from climatic covariates.

\begin{figure}[t]
\centering
  \begin{subfigure}[b]{.9\linewidth}
    \centering
\includegraphics[width=.99\textwidth]{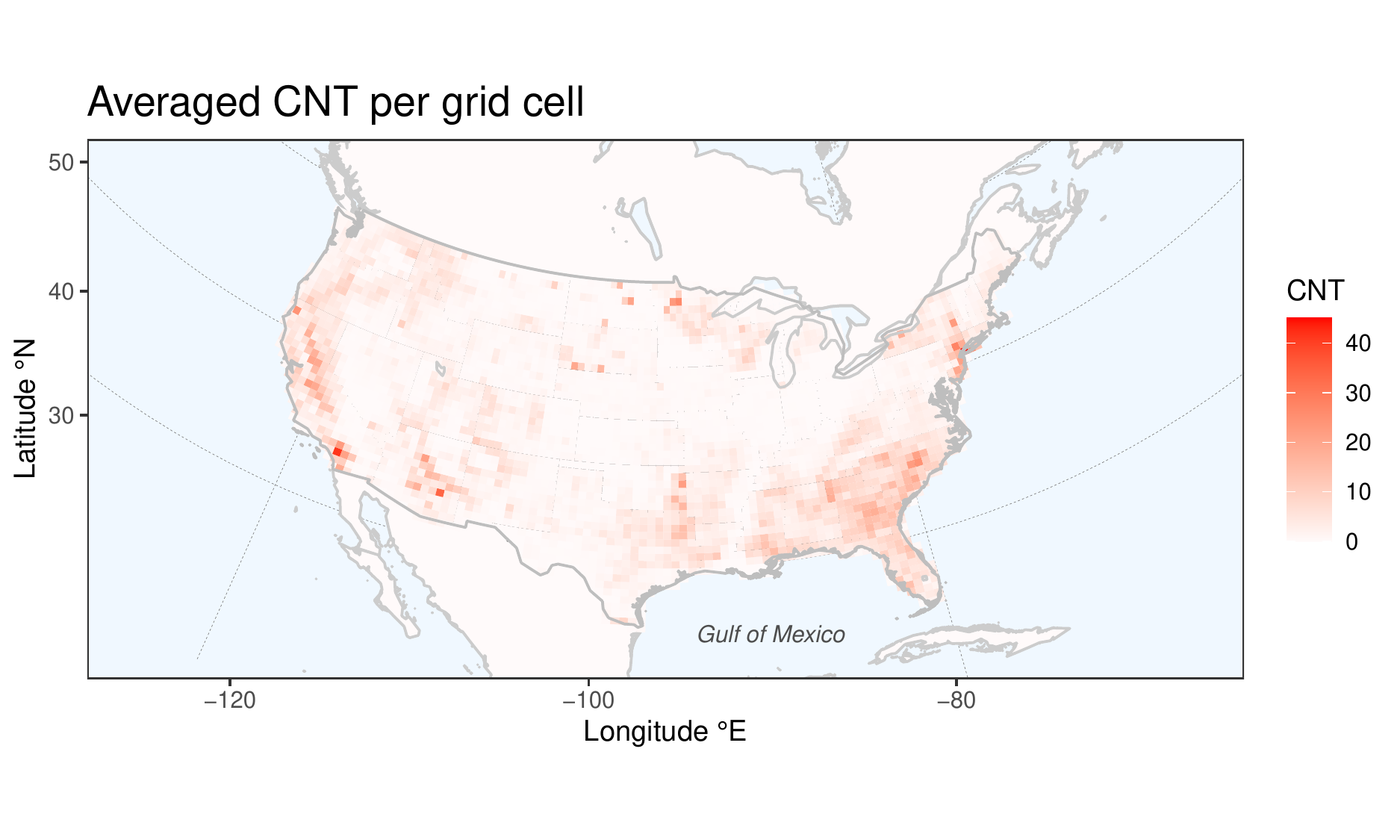}
  \end{subfigure}
  \begin{subfigure}[b]{.9\linewidth}
    \centering
    \includegraphics[width=.99\textwidth]{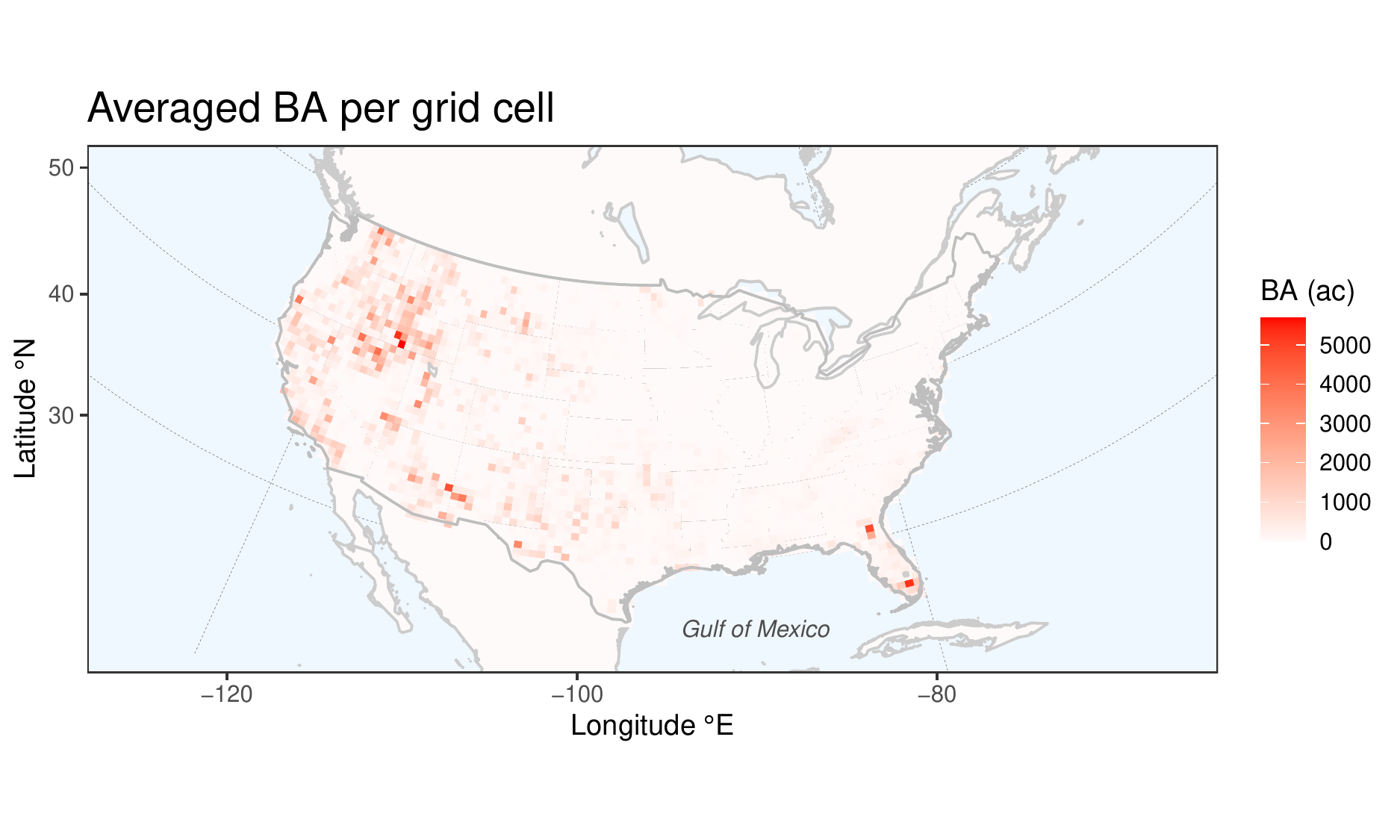}
  \end{subfigure} \\
  \caption{Maps of CNT (top) and BA (bottom) averaged across all months and years for each grid cell.}
  \label{fig:explore:CNTBA:spatial}
\end{figure}

The data comprise the monthly numbers of wildfires (CNT) and the aggregated burned area (BA) in each grid cell based on a $0.5\degree\times0.5\degree$ grid of longitude and latitude coordinates (roughly $55$km $\times 55$km) covering the study area, from March to September each year. 
Figure \ref{fig:explore:CNTBA:spatial} shows that the grid cells with the highest averaged CNT tend to be clustered towards the west (California) and southeast (North and South Carolina) corners of the study region, while clusters in the west (near the border of Idaho and Nevada), southwest (Arizona, New Mexico and Texas), and southeast (Florida) are observed for BA. 


\begin{figure}[t]
\centering
  \begin{subfigure}[b]{.48\linewidth}
    \centering
    \includegraphics[width=.99\textwidth]{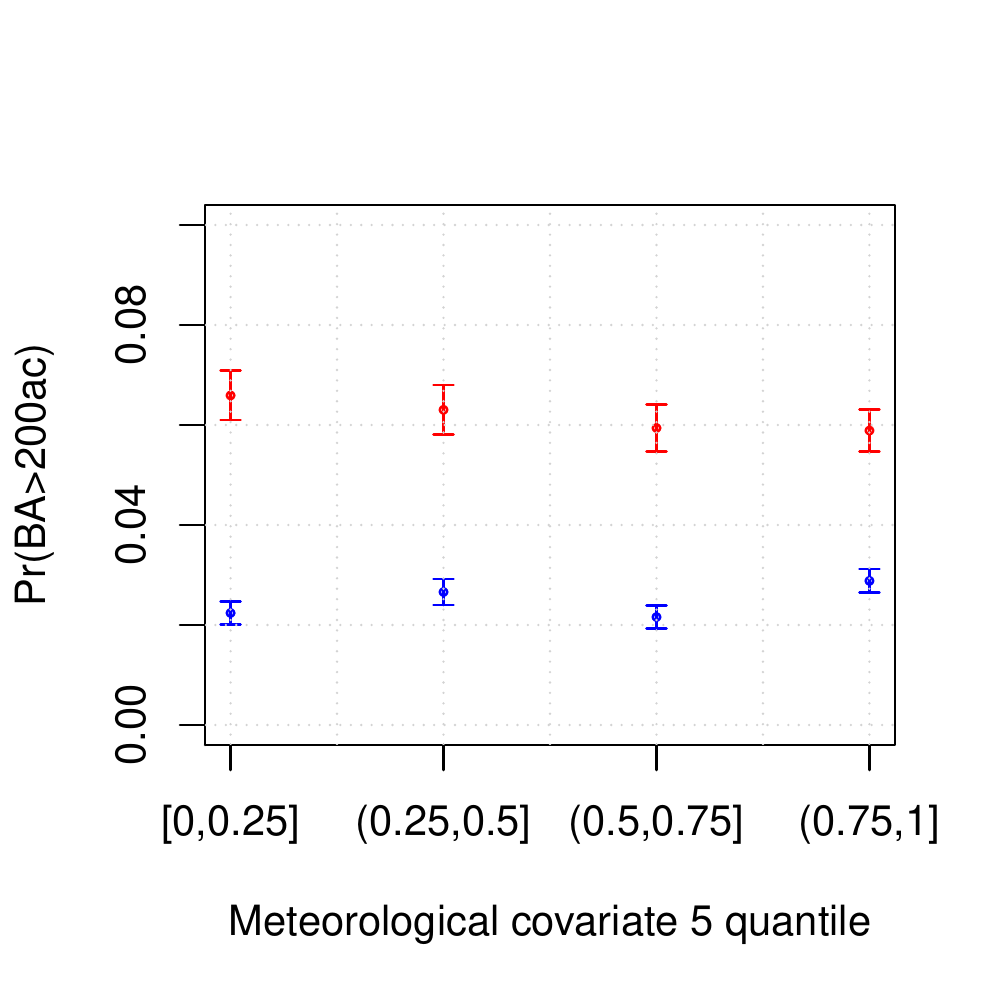}
  \end{subfigure}
  \begin{subfigure}[b]{.48\linewidth}
    \centering
    \includegraphics[width=.99\textwidth]{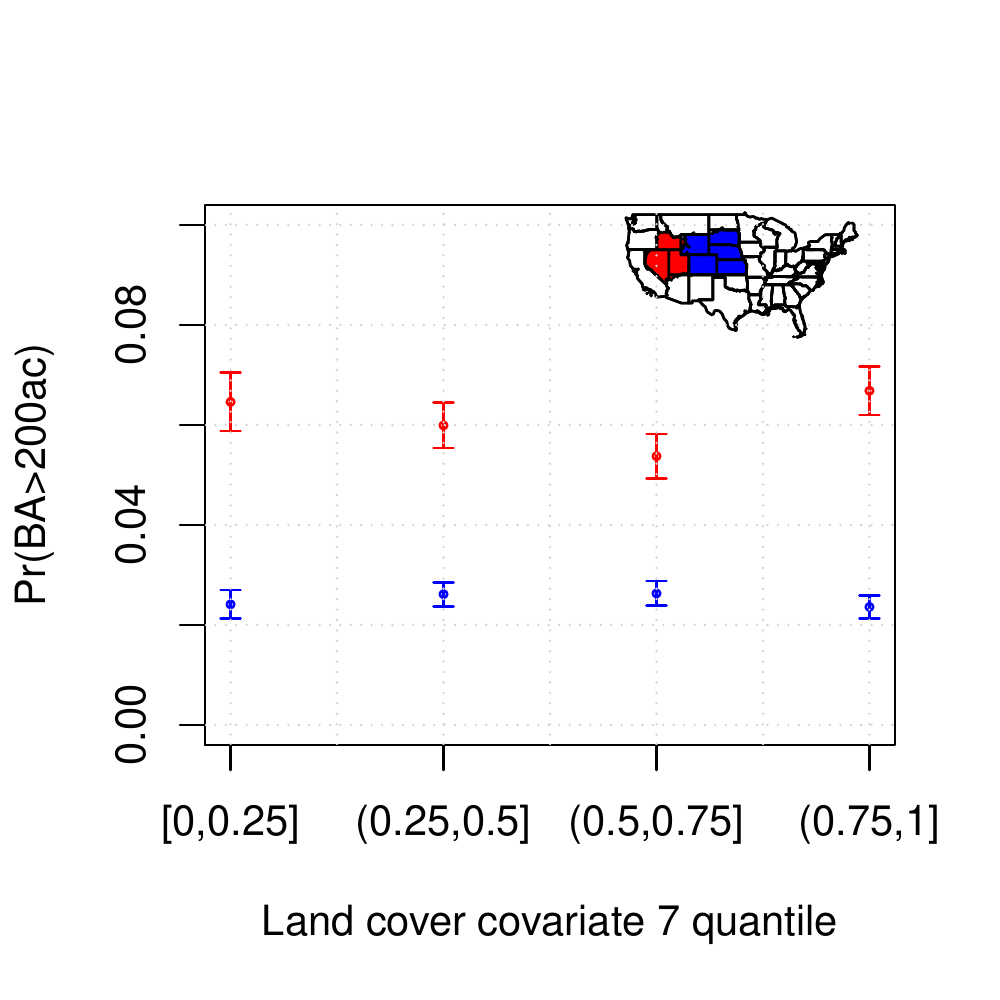}
  \end{subfigure} \\
  \caption{Empirical frequency $\text{Pr}(\text{BA}>200\text{ac})$ with $95\%$ error bars, as a function of the potential evaporation (in m, left) and tree needleleave evergreen closed to open (in $\%$, right) covariates, grouped by observations within four empirical quantile ranges, for grid cells within the Rocky Mountain Area (blue) and Great Basin (red) coordination centers. The two regions are highlighted in blue and red in the map inserted in the right panel. }   
  \label{fig:explore}
\end{figure}

Thirty-five auxiliary variables related to land cover, weather and altitude are provided at the same spatial and temporal resolution, and can be used for modelling. Figure \ref{fig:explore} hints at the importance of some of these variables, such as meteorological covariate 5 (clim5; potential evaporation, measured in meter water equivalent -- mwe) and land cover covariate 7 (lc7; tree needleleave evergreen closed to open, measured in $\%$ of the grid cell), for predicting high BA. However it also indicates that their effect differs over space, and that the interactions between these covariates may be important for predicting large burned areas. The Rocky Mountain Area and Great Basin, two regions defined formally as `Geographic Area Coordination Centers' by the United States Department of Agriculture, 
have empirical exceedance probabilities that respond differently to both covariates. For instance, the probability tends to decrease, then increase (after the $75\%$ quantile) with potential evaporation in the Great Basin, while the opposite holds for the Rocky Mountain Area, though the associated large uncertainties suggest that there is substantial heterogeneity within each region. 

\begin{figure}[t]
\centering
  \begin{subfigure}[b]{.45\linewidth}
    \centering
\includegraphics[width=.99\textwidth]{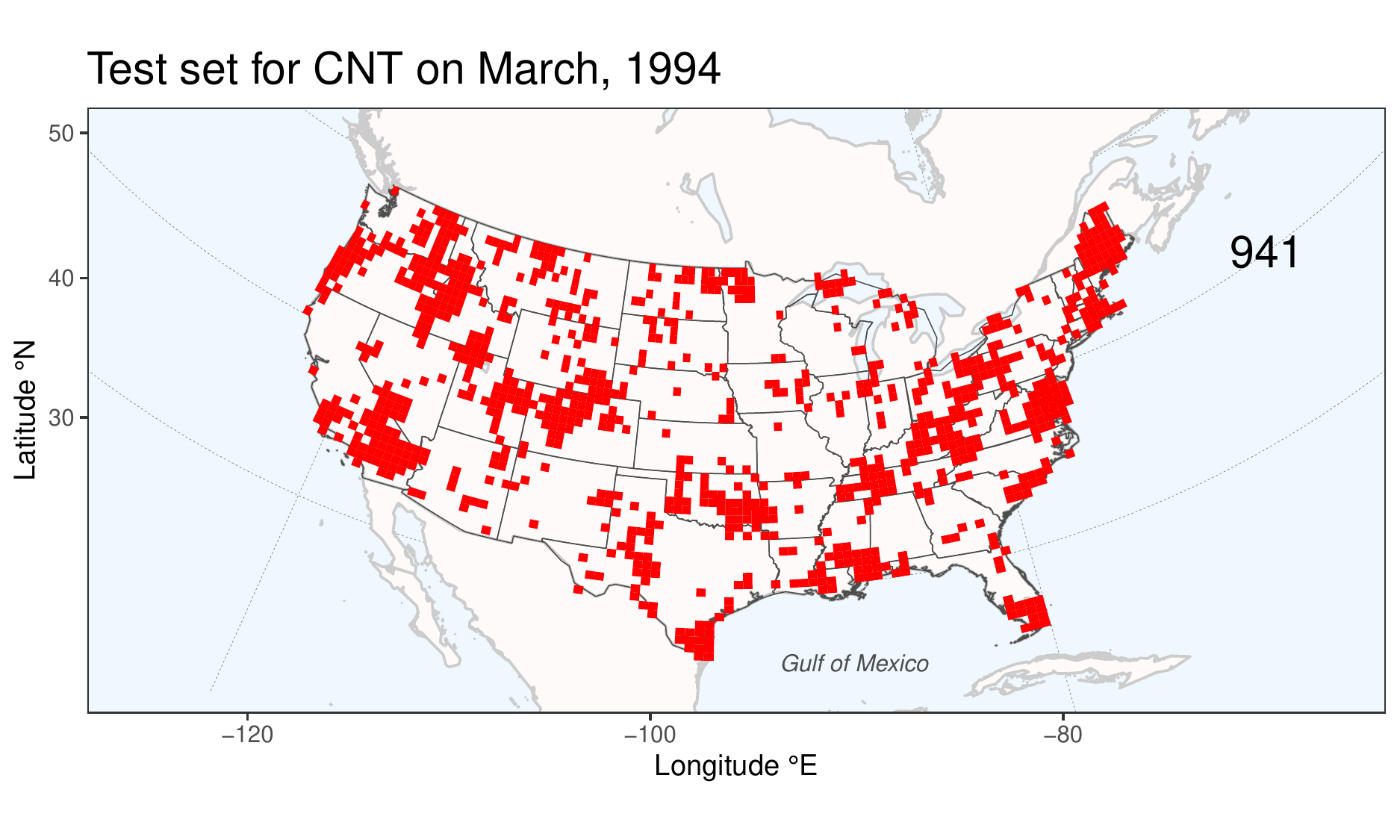}
  \end{subfigure}
  \begin{subfigure}[b]{.45\linewidth}
    \centering
    \includegraphics[width=.99\textwidth]{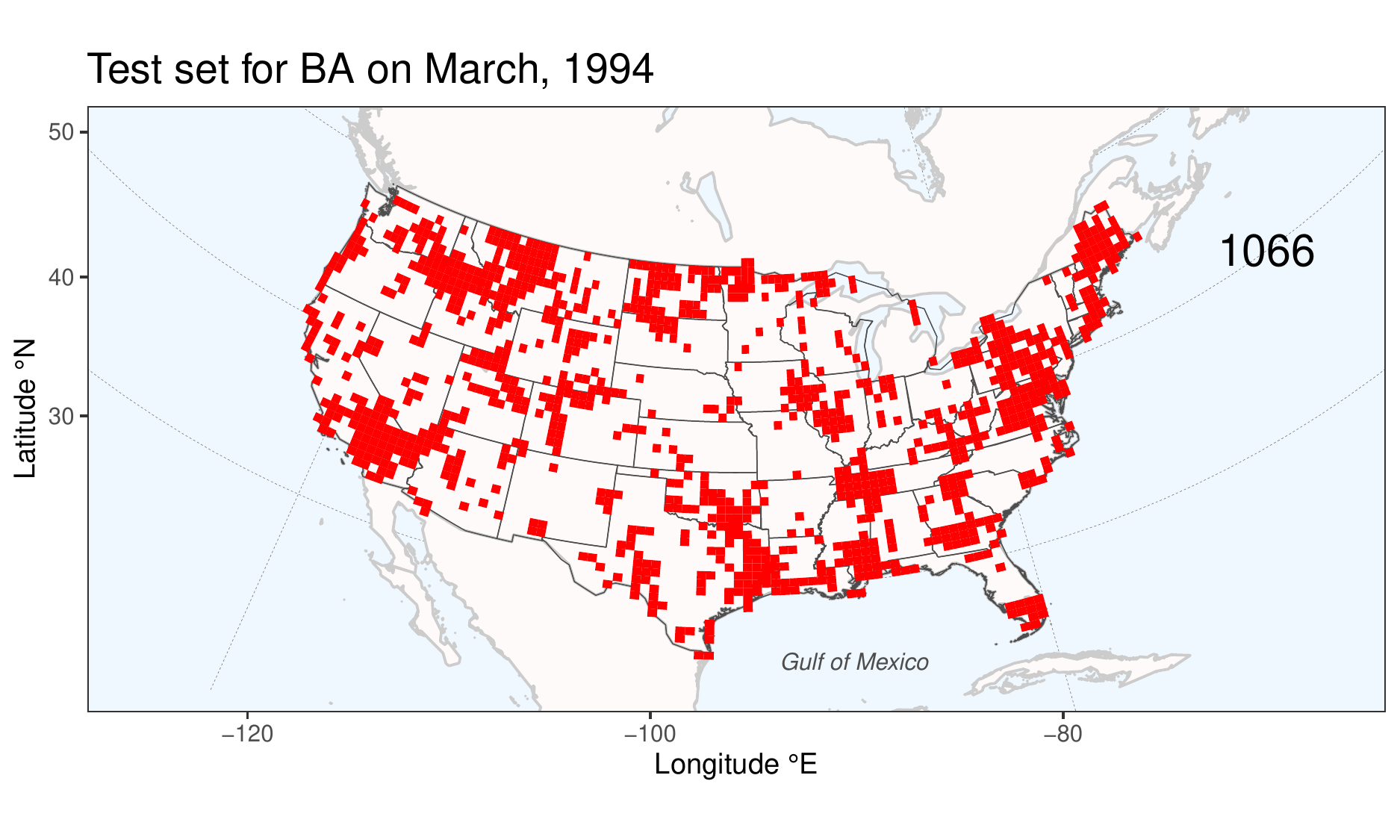}
  \end{subfigure} \\
  \caption{The test set grid cells (in red) for CNT (left) and BA (right) in March 1994. The number at the top right indicates the number of masked grid cells. }
  \label{fig:explore:validation_spatial}
\end{figure}

The original dataset has no gaps, but missing data were artificially created to compare predictive approaches; the full dataset was split into training and testing subsets to evaluate participants' test scores. No data were masked in the odd years, but 80,000 observations of each variable were masked in the even years. Figure \ref{fig:explore:validation_spatial} shows that the spatial and temporal positions of test data are clustered in space, and the test grid cells for BA and CNT are correlated. 
This masking is reminiscent of a real-world situation in which two related and spatially dependent processes could render both CNT and BA unobservable at small spatial clusters every month (e.g., from a potentially spatially dependent measurement error), and one could only use the available covariates and responses from the surrounding non-masked regions for prediction. \JON{For example, datasets generated using satellite-based remote sensing of wildfires have known misclassification issues, very often due to cloud occlusion.}

The evaluation metrics used for the competition \citep[see][]{data.challenge} require estimates of the probability $\text{Pr}(\text{BA}<u_{\text{BA}})$ and $\text{Pr}(\text{CNT}<u_{\text{CNT}})$ for $28$ thresholds $u_{\text{CNT}}$ and $u_{\text{BA}}$. The metrics are variants of weighted ranked probability scores and put relatively strong weight on good prediction in the extremes of the distributions of counts and burned areas. As such, we expect that models that emphasize accurate modelling of the largest counts and burned areas will perform better, and we achieve this by appealing to extreme-value theory.

\section{Methodology}\label{sec:methods-general_boosting}

\subsection{Extreme-value theory}

The generalized Pareto distribution (GPD) arises asymptotically for excesses above a large threshold of a random variable $X\sim F$, when the distribution $F$ satisfies mild regularity conditions. Let $x^\star=\sup\{x : F(x)<1\}$. The excess distribution above $u<x^\star$ can be approximated \citep{Davison-Smith.1990} as 
\begin{equation}\label{eq:gpd_boosting}
    \mathrm{Pr}(X>x+u \mid X>u) \approx 1-\mathrm{GPD}_{\sigma,\xi}(x) = \left \{
\begin{array}{ll}
(1+\xi x/\sigma)_{+}^{-1/\xi},& \quad \xi \neq 0, \\
\exp (-x/\sigma),& \quad \xi =0, 
\end{array}
\quad x >0,
\right.
\end{equation}
with shape parameter $\xi\in\mathbb{R}$ and  scale parameter $\sigma=\sigma(u)>0$, where $a_{+}=\max(a,0)$. 
The shape parameter determines the rate of tail decay, with slow power-law decay for $\xi>0$, exponential decay for $\xi=0$, and polynomial decay towards a finite $x^\star$, for $\xi<0$. When the approximation (\ref{eq:gpd_boosting}) is exact asymptotically (i.e., when $u \rightarrow x^\star$), we say that the random variable $X$ lies in the maximum domain of attraction of a generalized Pareto distribution with shape parameter $\xi$, written $X\in \text{MDA}_\xi$.

\subsubsection{Positive discrete random variables}

We say that a discrete non-negative random variable $Y$ lies in the discrete maximum domain of attraction, $Y\in \text{dMDA}_\xi$, if there exists a random variable $X\in \text{MDA}_\xi$ with $\xi\geq 0$ such that $\text{Pr}(Y\geq k) = \text{Pr}(X\geq k)$, for $k = 0,1,\dots$. The random variable $X$ is called the \textit{version} of $Y$, and many popular discrete distributions such as the geometric, Poisson and negative binomial distributions lie in $\text{dMDA}_\xi$. 

For $Y\in \text{dMDA}_\xi$ and large integers $u$,
\begin{align}\label{eq:dgpd}
     \text{Pr}(Y-u =k \mid Y\geq u) &\approx 
     \mathrm{GPD}_{\sigma,\xi}(k+1) - \mathrm{GPD}_{\sigma,\xi}(k) \nonumber \\ 
    &= (1+\xi k/\sigma)^{-1/\xi} - (1+\xi (k+1)/\sigma)^{-1/\xi},
\end{align}
where the last term is the probability mass function of the discrete generalized Pareto distribution \citep{shimura.2012}. Several studies have used this distribution to model count data; \citet{Prieto.2014} modelled numbers of road accidents and \citet{hitz.2017} modelled numbers of extreme tornadoes per outbreak and multiple births.

\subsection{Gradient tree boosting}\label{sec:model:boosting} 


The generic gradient boosting estimator \citep{Buhlmann.2007} is a sum of base procedure estimates. Regression trees \citep[CART]{Breiman.1984} are popular base procedures, as they include non-linear covariate interactions by construction, and are invariant under monotone transformations of covariates, so the user need not search for good data transformations. 

Let $\mathcal{D} = \{(\bm x_i, y_i) \}$, $\bm x_i \in \mathbb{R}^p $, $y_i \in \mathbb{R}$, $i=1,\dots,n$, be a dataset with $n$ observations and $p$ covariates. A \textit{binary split} of the covariate space uses a splitting variable indexed by $j\in \{1,\dots,p\}$ and a split point $v \in \mathbb{R}$ to partition the space into the pair of half-spaces $\{\bm x \in \mathbb{R}^p :  x_j \leq v\}$ and  $\{\bm x \in \mathbb{R}^p :  x_j > v\}$, where $x_j$ is the $j$-th component of $\bm x$.

By successive binary splits, a regression tree partitions the covariate space into a set of $L$ disjoint regions $A_1,\dots,A_L$, and fits a simple model such as a constant in each region.
The regions created by the splits are called nodes; a terminal node is called a \textit{leaf} and an interior node is called a \textit{branch}. We index leaves by $l\in\{1,\dots,L\}$, with leaf $l$ representing region $A_l$. The simplest tree is one with two leaves, known as a stump. A learning algorithm needs to decide the tree structure, i.e., the splitting variables and split points. 

Suppose that $L$ leaves with regions $A_1,\dots, A_L $ have been chosen and we model the response as a \textit{score} $c_l \in \mathbb{R}$ in each region. A regression tree is a function
$$
f(\bm x_i) = \sum^{L}_{l=1} c_l \mathbb{I}(\bm x_i \in A_l),
$$
where $\mathbb{I}$ is the indicator function. A gradient tree boosting model uses $T$ such trees to model the \textit{boosting estimate} 
\JON{\begin{equation}\label{eq:boosting_estimate}
    \hat{\theta}_i = \sum_{t=1}^{T} f_t(\bm x_i) , \quad f_t \in \mathcal{T},
\end{equation}
}where $\mathcal{T}$ is the space of regression trees. \JON{In this paper, $\hat{\theta}_i$ will always represent a parameter estimate in a given model for the conditional distribution of $y_i$ given $\bm x_i$.} The boosting estimate using stumps will be additive in the original covariates, because every base estimate is a function of a single covariate. A boosting model that has trees with at most $L$ leaves has interactions of order at most $L-2$. Thus, constraining the maximum number of nodes in the base procedure controls the complexity of the model.


Gradient tree boosting learns the set of trees used in (\ref{eq:boosting_estimate}) by minimizing a regularized objective function in a greedy iterative fashion; at each iteration we add the tree that most improves our model according to an objective function $\mathcal{O}$. More precisely, let $\hat{\theta}_i^{(j)}$ be the boosting estimate for the $i$-th observation at the $j$-th iteration. We add a tree $f_j$ at each iteration to minimize 
$$
\mathcal{O}^{(j)} = \sum_{i=1}^n \mathcal{L}\{y_i, \hat{\theta}_i^{(j-1)} + f_j(\bm x_i)\} + \Omega(f_j), 
$$
where $\Omega(f_j) = \eta L^{(j)} + \lambda ||\bm c||^2/2$, $\mathcal{L}$ is a differentiable loss function, $L^{(j)}$ is the number of leaves in the tree $f_j$ and $\bm c\in \mathbb{R}^{L^{(j)}}$ is the corresponding vector of scores. 
The regularization term $\Omega$ is added to penalize the complexity of each tree, and the positive parameters $\eta$ and $\lambda$ control the penalization. The form of $\Omega$ is simple enough to allow parallel \JON{computation} \citep{Chen.2016}.

Using a second-order approximation for the objective \citep{Friedman.etal.2000} gives
\begin{equation}\label{eq:objective}
\mathcal{O}^{(j)} \simeq \sum_{i=1}^n {\left\{ \mathcal{L}(y_i, \hat{\theta}_i^{(j-1)} ) + g_i f_j(\bm x_i) + h_i f_j^2(\bm x_i) \right\}} + \Omega(f_j), 
\end{equation}
where $g_i= \partial \mathcal{L} (y_i, \hat{\theta}_i^{(j-1)} )/ \partial \hat{\theta}_i^{(j-1)}$ and $h_i= \partial^2 \mathcal{L}(y_i, z_i )/ \partial z_i^2 \mid_{z_i=\hat{\theta}_i^{(j-1)}}$. We then minimize (\ref{eq:objective}) with respect to the tree structure and weight vector $\bm c$.

Let $I_l =\{i : \bm x_i \in A_l \}$ denote the \textit{instance set} of leaf $l$. For a fixed tree structure with regions $A_1, \dots, A_{L^{(j)}}$, the optimal weights $\bm c^\star$ can easily be found and have components
$$
c^\star_l = -\dfrac{\sum_{i\in I_l} g_i }{\sum_{i\in I_l} h_i + \lambda}, \quad l=1,\dots, L^{(j)}.
$$
Plugging the weights $\bm c^\star$ into (\ref{eq:objective}) and removing the term that does not depend on $f_j$ gives
\begin{equation}\label{eq:impurity}
\tilde{\mathcal{O}}^{(j)} = -\dfrac{1}{2}\sum_{l=1}^{L^{(j)}} \dfrac{ (\sum_{i\in I_l} g_i )^2 }{\sum_{i\in I_l} h_i + \lambda} + \eta L^{(j)},
\end{equation}

which can be used as a scoring function to measure the quality of the tree structure, a role similar to the impurity score in \citet{Breiman.1984}. 

Assume that a split has been performed, and let $I_L$ and $I_R$ denote the instance sets of the left and right leaves from this split. Define $I=I_L \cup I_R $. The loss reduction from this split is
\begin{equation}\label{eq:split}
G = \dfrac{1}{2} \Bigg\{ \dfrac{ (\sum_{i\in I_L} g_i )^2 }{\sum_{i\in I_L} h_i + \lambda} + \dfrac{ (\sum_{i\in I_R} g_i )^2 }{\sum_{i\in I_R} h_i + \lambda} - \dfrac{ (\sum_{i\in I} g_i )^2 }{\sum_{i\in I} h_i + \lambda} \Bigg\} - \eta ,
\end{equation}
and (\ref{eq:split}) is used for evaluating candidate split variables and points.

As it is impossible to enumerate all possible tree structures, most existing tree boosting implementations, such as in \texttt{scikit-learn} \citep{Pedregosa.sklearn.2011} and \texttt{gbm} \citep{greenwell.gbm}, use greedy algorithms that start from a single leaf and iteratively add branches to the tree based on (\ref{eq:split}), until the gain for the best split is negative. Here we use the greedy algorithm implemented in \texttt{xgboost} \citep[called the approximate algorithm with weighted quantile sketch][Appendix A]{Chen.2016}, 
that further reduces computational cost and parallelizes computations when the data do not fit into memory. This algorithm proposes candidate splitting points from the empirical quantiles of each covariate, instead of considering all possible splitting points for each variable. We also subsample a proportion $s\in[0,1]$ of the covariates at each iteration, like in the random forest algorithm \citep{Breiman.2001}, to further prevent overfitting and to accelerate parallel computation. \JON{For the full algorithmic details, we refer the reader to \citet{Chen.2016}.} 

\JON{Convexity of $\mathcal{L}$ is a desirable feature that would guarantee that a unique global minimum exists; if this does not hold in practice and one is concerned about the algorithm being stuck at local minimums, then a potential solution is to rerun the algorithm multiple times from different initial boosting estimates $\hat{\theta}_i^{(0)}$, $i=1,\dots,n$, and select the run that gives the lowest loss.} \JON{If $\mathcal{L}$ is convex and there are enough iterations, then the choice of the initial boosting estimate will have minimal effect on the overall prediction. A natural choice is to set $\hat{\theta}_i^{(0)}=\hat{\theta}_j^{(0)}=\hat{\theta}$, for all $i\neq j$, where $\hat{\theta}= \arg \min_\theta \sum_{i=1}^n \mathcal{L}(y_i, {\theta})$, i.e., this is the common parameter value that minimises the loss across all observations, unconditionally on the covariates.}






As gradient tree boosting is an ensemble method combining predictions from many trees, the exact relationship between covariates (or their interactions) and the response is difficult to determine. 
Suppose we treat the covariates as random, and let $\bm X_\mathcal{S}$ denote the random subvector (of size $l<p$) of the full covariate vector $\bm X = (X_1, \dots, X_p)^T $, indexed by the set $\mathcal{S} \subset \{1,\dots,p\}$. Let $\mathcal{C}$ be the set complementary to $\mathcal{S}$. For a predictive function $f$ at a fixed point $\bm x \in \mathbb{R}^l$, the partial dependence function \citep{friedman.2001},
$$
f_\mathcal{S}(\bm x) =  \text{E}_{\bm X_\mathcal{C}}  f\{ (\bm x, \bm X_\mathcal{C})^T \} ,
$$
can be estimated by Monte Carlo as
\begin{equation}\label{eq:pdp}
    \hat{f}_\mathcal{S}(\bm x) =  \dfrac{1}{n} \sum_{i=1}^n f\{ (\bm x, \bm x_{i\mathcal{C}})^T \},
\end{equation}
where $\bm x_{1\mathcal{C}}, \dots, \bm x_{n\mathcal{C}} $ are realisations of $\bm X_{\mathcal{C}}$ from the training data. 

Metrics used to rank covariates in terms of their importance include the \textit{coverage} for a chosen covariate, which is the sum of the second order gradients $h_i$ from (\ref{eq:impurity}), in each node which uses this covariate, standardized by dividing by the sum of the metrics for all other covariates (so the resulting metric is a proportion). The \textit{gain} metric represents the fractional contribution of a chosen covariate to the model based on the total gain of all the splits involving this covariate, measured by $G$ in (\ref{eq:split}); it is the total improvement of the model in terms of the objective function, from the branches that include the covariate. In both cases a higher proportion implies a more important predictive variable.

The loss function $\mathcal{L}$ in (\ref{eq:objective}) strongly governs the type of models that we fit, and we discuss choices for this next.

\subsection{Loss functions}\label{sec:lossfunctions}

\subsubsection{For wildfire counts}\label{sec:loss:counts}
  
\JON{The most common statistical model for count data is the Poisson distribution, and many existing boosting implementations use a scaled version of the simplified Poisson loss as their default loss function for counts.} Given $n$ monthly wildfire counts in a grid cell, $y_1, \dots, y_n$, this loss is
\begin{equation}\label{eq:loss:poisson}
\mathcal{L}_\text{Pois}(y_i, \hat{\theta}_i) =   y_i \log\{y_i/\exp(\hat{\theta}_i)\} - y_i+\exp(\hat{\theta}_i) ,
\end{equation}
where Stirling's approximation $\log(y_i!) \approx y_i \log(y_i) - y_i$ is used and the boosting estimate $\hat{\theta}_i$ models the log mean of the $i$-th Poisson count. \JON{The terms that do not depend on $\hat{\theta}_i$ can be dropped when minimising this loss in an optimisation procedure.} Although (\ref{eq:loss:poisson}) is also the most common choice for modelling wildfire counts in the literature, the zero-inflation and potentially heavy tails of count distributions suggest that it may be unsuitable.

Instead, if we let $\alpha=1/\xi>0$ and $\lambda=\sigma \alpha$ in (\ref{eq:dgpd}), we motivate a new loss function for counts from extreme-value theory, the discrete generalized Pareto (dGPD) loss
\begin{equation}\label{eq:loss:dpgd}
    \mathcal{L}_\text{dGPD}(y_i, \hat{\theta}_i) = \{1+\exp(\hat{\theta}_i) y_i\}^{-\alpha} - \{1+\exp(\hat{\theta}_i)(y_i+1)\}^{-\alpha}.
\end{equation}
The boosting estimate $\hat{\theta}_i$ models the logarithm of the rescaled scale parameter $\lambda_i$. If $\alpha>1$, the predicted mean of the $i$-th count is 
\begin{equation}\label{eq:dgpd:mean}
   \hat{m}_i= \sum_k^{\infty} 1/ \{1+ \exp(\hat{\theta}_i) k \}^\alpha ,
\end{equation}
and otherwise the mean does not exist.

The first and second derivatives of (\ref{eq:loss:dpgd}) with respect to the boosting estimate, i.e.,
\begin{align*}
g^\text{dGPD}_i=  & -\alpha \{ 1 + \exp(\hat{\theta}_i)y_i \}^{-\alpha-1} \{ \exp(\hat{\theta}_i) y_i \} \\ &+ \alpha \{ 1 + \exp(\hat{\theta}_i)(y_i+1) \}^{-\alpha-1} \{ \exp(\hat{\theta}_i) (y_i+1) \}, \\
h^\text{dGPD}_i=  & -\alpha (-\alpha-1) \{ 1 + \exp(\hat{\theta}_i)y_i \}^{-\alpha-2} \{ \exp(\hat{\theta}_i) y_i \}^2 \\ &- \alpha \{ 1 + \exp(\hat{\theta}_i)y_i \}^{-\alpha-1} \{ \exp(\hat{\theta}_i) y \} \\ &+ \alpha (-\alpha-1) \{ 1 + \exp(\hat{\theta}_i)(y_i+1) \}^{-\alpha-2} \{ \exp(\hat{\theta}_i) (y_i+1) \}^2 \\ &+ \alpha \{ 1 + \exp(\hat{\theta}_i)(y_i+1) \}^{-\alpha-1} \{ \exp(\hat{\theta}_i) (y_i+1) \},
\end{align*}
are used in the second-order approximation of the objective in (\ref{eq:objective}) and are essential for determining split variable and split point candidates when building trees with (\ref{eq:split}).

\subsubsection{For wildfire sizes}\label{sec:loss:size}

The squared loss is the dominant choice in the literature on predicting burned areas in this regression context, but it implicitly presupposes normality of the response conditional on the covariates, which may be inappropriate if the distributional tail decays more slowly than exponential. Moreover, burned areas cannot be negative. Modelling log-transformed burned areas addresses the latter issue, though doing so still excludes conditional distributions with Pareto-like tails.

Another approach to modelling the full distribution of wildfire sizes is to use a mixture, by first choosing an appropriately high threshold and then fitting the burned areas below and above that threshold with different loss functions. \JON{To additionally handle the zero-inflation of wildfire sizes, we can left truncate the sizes below the high threshold at zero. This mixture approach models burned areas by splitting the distribution into three groups: zero, intermediate and extreme sizes.}

To model the monthly burned area in a grid cell, $y_i$, below a chosen threshold $u>0$, the negative log-loss likelihood of a truncated distribution could be used. The probability density function of a right truncated gamma distribution is
\begin{equation*}
   f(x) = \left \{
\begin{array}{ll}
 \dfrac{(\mu/k)^\JON{-k} x^{k-1}\exp(-xk/\mu) }{ \gamma(k, k u/\mu) } &,\quad 0<x\leq u, \\
\JON{0} &,\quad x>u, 
\end{array}
\right.
\end{equation*}
where $u>0$ is the right truncation, $\mu>0$ is the rescaled scale parameter, $k>0$ is the shape parameter and $\gamma(k,s) = \int_0^{s} t^{k-1} \exp(-t) \text{d}t$, $s>0$, is the lower incomplete gamma function. Modelling $\log(\mu)$ with the boosting estimate \JON{and dropping the terms that do not depend on $\hat{\theta}_i$} gives
\begin{align*}\label{eq:loss:gamma}
    \mathcal{L}_\text{trGamma}(y_i, \hat{\theta}_i) = &\JON{k\hat{\theta}_i} 
     +y_ik/\exp(\hat{\theta}_i) + \log \gamma\{k, k u/\exp(\hat{\theta}_i)\}, \\
g^\text{trGamma}_i=  & \exp(\hat{\theta}_i) \{ k/\exp(\hat{\theta}_i) - y_ik/\exp(\hat{\theta}_i)^2 \\ & +\gamma'\{k, k u/\exp(\hat{\theta}_i)\}/\gamma\{k, k u/\exp(\hat{\theta}_i)\} \} , \\
h^\text{trGamma}_i=  & \exp(\hat{\theta}_i)^2 \bigg\{ -k/\exp(\hat{\theta}_i)^2 + 2y_ik/\exp(\hat{\theta}_i)^3 \\ &+ \dfrac{ \gamma\{k, k u/\exp(\hat{\theta}_i)\} \gamma''\{k, k u/\exp(\hat{\theta}_i)\} - \gamma'\{k, k u/\exp(\hat{\theta}_i)\}^2 }{ \gamma\{k, k u/\exp(\hat{\theta}_i)\}^2 } \bigg\} \\ &+
g^\text{trGamma}_i ,
\end{align*}
where  $\gamma'\{k, k u/\exp(\hat{\theta}_i)\}$ and $\gamma''\{k, k u/\exp(\hat{\theta}_i)\}$ are the first and second derivatives of $\gamma\{k, k u/\exp(\hat{\theta}_i)\}$ with respect to $\hat{\theta}_i$, and have closed forms (see Supplement \S\ref{sec:appendix}). 

To model only the excesses above a threshold $u$, we can use the GPD in (\ref{eq:gpd_boosting}), and assume that $\xi>0$, since burned areas tend to be heavy-tailed. 
If we reparameterize and model the logarithmic $\kappa \in [0,1]$ quantile of the excesses with the boosting estimate, i.e., $\hat{\theta}_i = \log[ \{(1-\kappa)^{-\xi}-1\}\sigma_i  / \xi ] $, we can define the generalized Pareto (GPD) loss and obtain its derivatives 
\begin{align*}
 \mathcal{L}_\text{GPD}(y_i, \hat{\theta}_i) = & \dfrac{\xi+1}{\xi} \log\left[1+ \dfrac{y_i \{(1-\kappa)^{-\xi}-1\} }{  \exp(\hat{\theta}_i)} \right] + \JON{\log \left[\dfrac{\xi \exp(\hat{\theta}_i)}{\{(1-\kappa)^\xi - 1\} } \right] } , \\
g^\text{GPD}_i=  & -\dfrac{f'\{y_i, \exp(\hat{\theta}_i), \xi \}   }{ f\{y, \exp(\hat{\theta}_i), \xi\} }, \\
h^\text{GPD}_i=  & -\dfrac{f\{y_i, \exp(\hat{\theta}_i), \xi\} f''\{y_i, \exp(\hat{\theta}_i), \xi\} - f'\{y_i, \exp(\hat{\theta}_i\}, \xi\}^2  }{f\{y_i, \exp(\hat{\theta}_i), \xi\}^2} 
\end{align*}
where $f'$ and $f''$ are the first and second derivatives of the reparameterized probability density function $f\{y_i,\exp(\hat{\theta}_i), \xi\}$ given in Supplement \S\ref{sec:appendix}. 

  
  

\subsubsection{For wildfire size classification}\label{sec:size:classification}

Adopting the mixture modelling approach to wildfire sizes requires us to model the probability that a fire belongs to each size component $1,\dots,C$, where $C$ is the number of components. Let $\bm y_i= (y_{i,1}, \dots, y_{i,C})$ denote the vector of wildfire size component indicators, where $y_{i,c}=1$ if the $i$-th fire size is in component $c$, and otherwise $y_{i,c}=0$. We can model the probability of each class with the boosting estimate using the softmax function $\sigma: \mathbb{R}^{C} \rightarrow [0,1]^C$ defined by
\begin{equation}\label{eq:softmax}
\sigma(\bm z)_i = \dfrac{\exp(z_i)}{\sum_{j=1}^C \exp(z_j) } , \quad i=1,\dots,C, \quad \bm z=(z_1,\dots,z_C)\in\mathbb{R}^C.
\end{equation} 
The generalization of the logistic \citep{cox.1958} loss to multiple classes is the cross-entropy loss, which can be reweighted to give
\begin{equation}\label{eq:cross_entropy}
    \mathcal{L}_\text{CE}(\bm y_i, \hat{\bm \theta}_i) = -w_{i} \sum_{c=1}^C   y_{i,c} \log \Big\{ \exp(\hat{\theta}_{i,c})/ \sum_{d=1}^C \exp(\hat{\theta}_{i,d})  \Big\} ,
\end{equation}
where the vector of component probabilities is modelled by applying (\ref{eq:softmax}) to the boosting estimate $\hat{\bm \theta}_i = (\hat{\theta}_{i,1},\dots,\hat{\theta}_{i,C})^{T} $ , and the weights $w_1,\dots,w_n$ could be chosen to improve predictions in unbalanced classification tasks.

\subsection{A spatiotemporal cross-validation scheme }\label{sec:methods:cv}

The use of $k$-fold cross-validation generally presupposes independent replicates, so it would produce optimistic predictive performance estimates in our setting because data points that are geographically closer will have stronger dependencies. To address this, we first study the spatiotemporal process leading to grid cells being masked, which we call the \textit{masking process}. Figure \ref{fig:explore:validation_spatial} hints at either a common or two inter-correlated spatially dependent latent processes governing the observed masking processes for CNT and BA, which we model with a common latent Gaussian process \citep{Rasmussen.2005}. We then fit a Bernoulli model to observations arising from the masking process, and simulate observations from the model to generate cross-validation folds.


We consider only the months $m$ with masked observations, and let $M$ denote the number of those months and $D$ denote the number of grid cells. Let $R^\mathrm{CNT}_{d,m}$ and $R^\mathrm{BA}_{d,m}$ denote the binary 0-1 observations indicating whether the grid cell $d\in \{1,\dots,D\}$ (with centroid $\bm s_d$) at month $m\in \{1,\dots,M\}$ was masked for the CNT and BA responses, respectively. Our hierarchical model for the masking processes is
\begin{align*}
R^\mathrm{CNT}_{d,m}\mid \mu^{\mathrm{CNT}}_{dm} \sim\  &\mathrm{Bernoulli}\{ \text{expit}( \mu^{\mathrm{CNT}}_{dm} )\}, \\
R^\mathrm{BA}_{d,m}\mid \mu^{\mathrm{BA}}_{dm} \sim\  &\mathrm{Bernoulli}\{ \text{expit}( \mu^{\mathrm{BA}}_{dm} )\};
\end{align*}
\begin{align*}
\mu^\mathrm{CNT}_{dm} =  &\beta_0^\mathrm{CNT} + { g^\text{}_m ( \bm s_d ) } + \epsilon^\text{CNT}_m, \\
\mu^\mathrm{BA}_{dm} =  &\beta_0^\mathrm{BA} + \beta { g^\text{}_m ( \bm s_d ) } + \epsilon^\text{BA}_m; 
\end{align*}
\begin{align*}
   g_m(\bigcdot) \sim &\mathcal{GP}(\bm \zeta), \\
      \epsilon^\text{CNT}_m,  \epsilon^\text{BA}_m \sim &\mathcal{N}( 0, \phi), \\
      \beta \sim &\mathcal{N}( 0, \omega) ;
\end{align*}
\begin{align*}
   \bm \zeta = \{\beta_0, \beta, \bm \zeta , \phi, \omega\} \sim \text{Priors},
\end{align*}
where $\text{expit}(x) = \{1+\exp(-x)\}^{-1}$ is the inverse logit function.

We fit this model using the integrated nested Laplace approximation \citep[INLA,][]{Rue.al.2009}, which is an approximate Bayesian inference technique well-suited for latent Gaussian models. The parameter $\beta$ governs the degree of latent sharing between the two masking processes and we use a  flat and independent zero-centered Gaussian hyperprior for it. Similar frameworks were used by \citet{Koh.2021} for the joint modelling of different wildfire risk components and by \citet{Diggle2010} and \citet{Pati.2011} to model preferential sampling. The spatial process $g_t$ is independently replicated in time and each replicate has a Gaussian process prior $\mathcal{GP}$ with a Mat\'ern covariance structure governed by the parameter vector $\bm \zeta$. We represent these Gaussian processes via a numerically convenient Gauss--Markov random field approximation, constructed by solving a stochastic partial differential equation \citep{Lindgren.al.2011}. Supplement \S\ref{sec:appendix} details the full procedure. 

\begin{figure}[t]
\centering
  \begin{subfigure}[b]{.45\linewidth}
    \centering
\includegraphics[width=.99\textwidth]{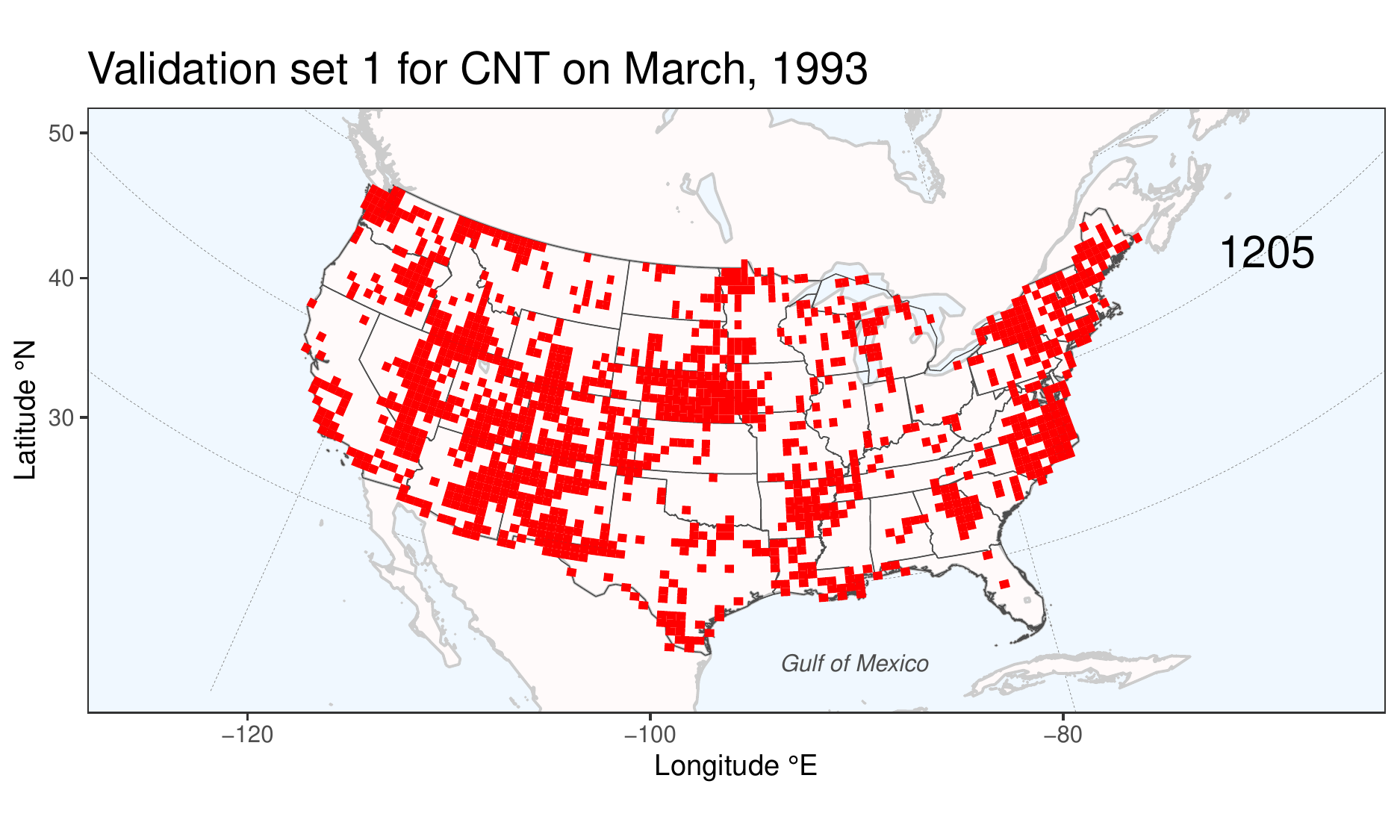}
  \end{subfigure}
  \begin{subfigure}[b]{.45\linewidth}
    \centering
    \includegraphics[width=.99\textwidth]{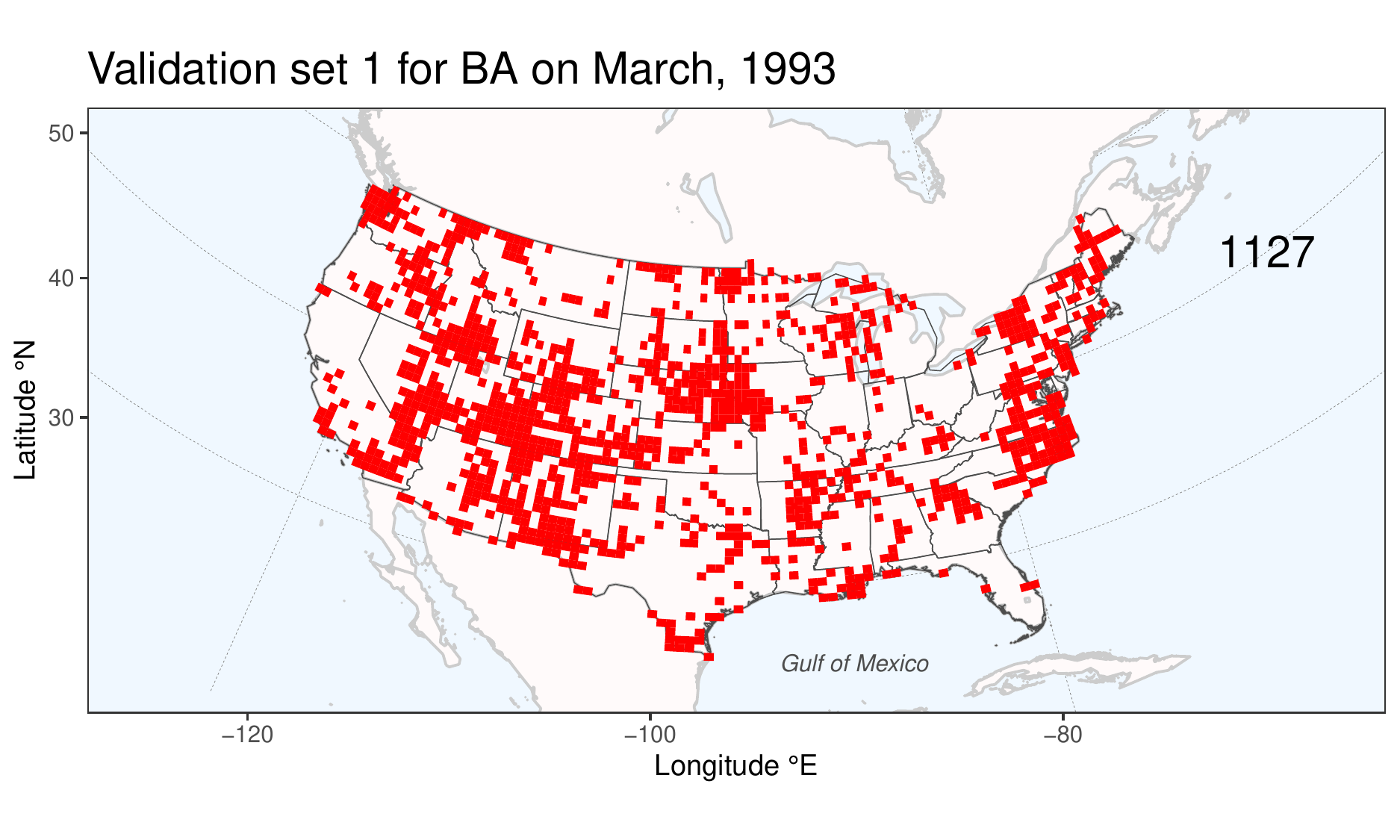}
  \end{subfigure} \\
    \begin{subfigure}[b]{.45\linewidth}
    \centering
\includegraphics[width=.99\textwidth]{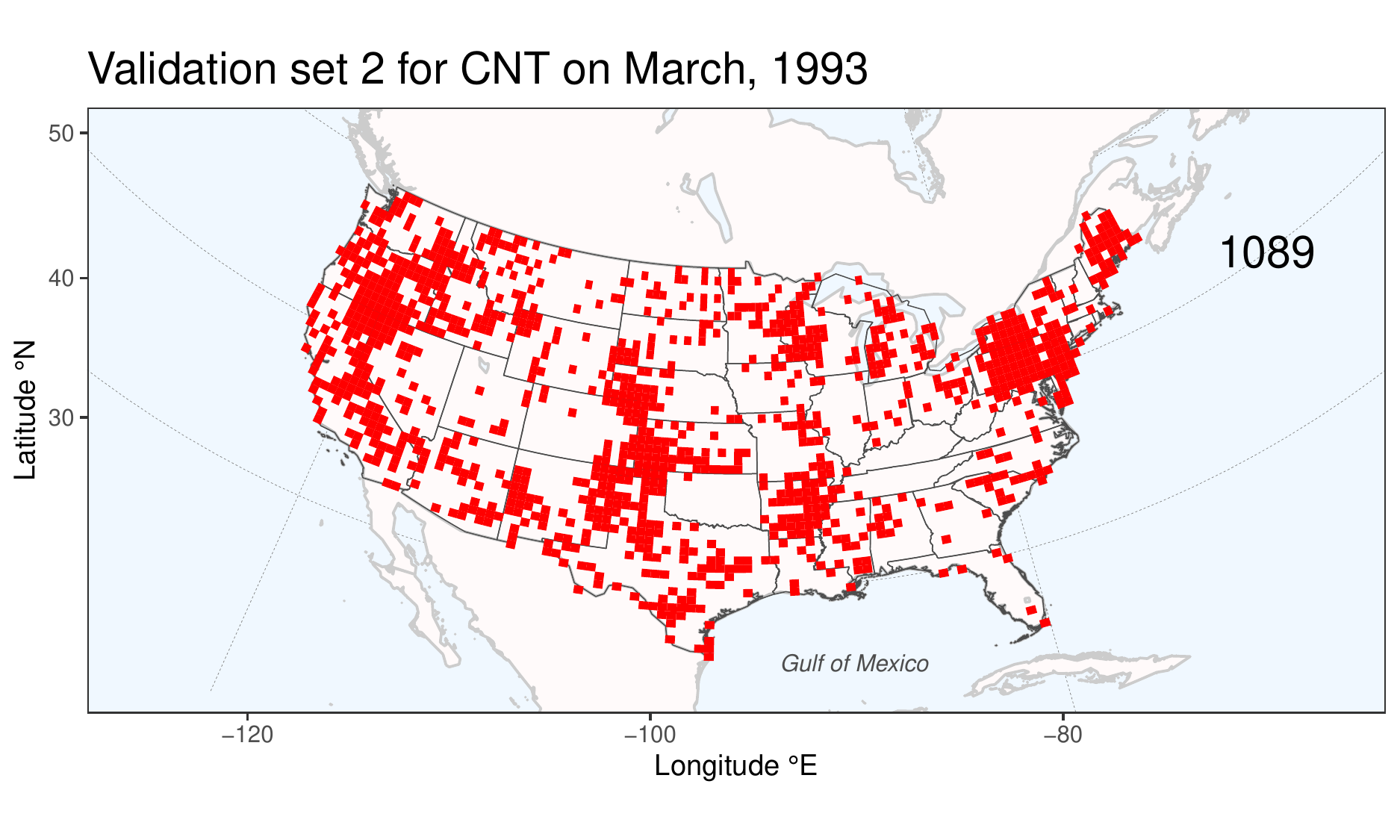}
  \end{subfigure}
  \begin{subfigure}[b]{.45\linewidth}
    \centering
    \includegraphics[width=.99\textwidth]{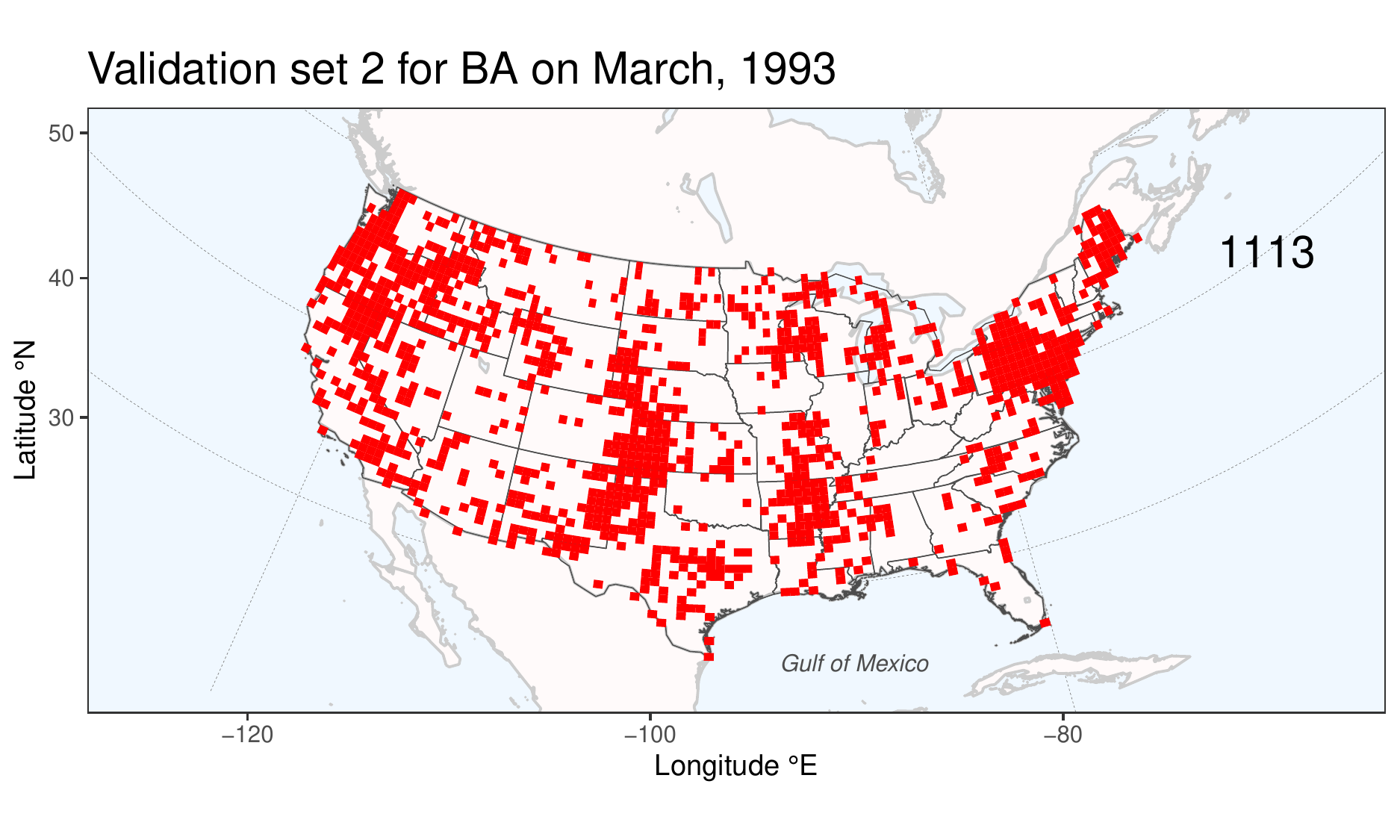}
  \end{subfigure} 
  \caption{The first (top) and second (bottom) validation folds (in red) for CNT (left) and BA (right) for March, 1993, from our spatiotemporal cross-validation scheme, generated from the Bayesian spatial model. The number at the top right indicates the sum of grid cells chosen. }
  \label{fig:spatial_cv_fold}
\end{figure}

We generate samples from this Bayesian model by first sampling parameters from the posterior distribution, and then generating observations from the Bernoulli model with the sampled parameters. We do this for all months, including in those where observations were masked; if a location was already part of the test set, i.e., if it was already masked, then we removed it from the validation set. Five samples were generated to obtain five folds for our cross-validation scheme. Figure \ref{fig:spatial_cv_fold} shows two samples from this model for March $1993$. 
The degree of spatial and inter-variable dependencies resemble those of the masking processes in Figure \ref{fig:explore:validation_spatial}, and the numbers of grid cells masked and chosen for validation in each month are also similar. The triplet ($2.5\%$ posterior quantile, posterior mean, $97.5\%$ posterior quantile) for the scaling parameter $\beta$ is $(0.28,0.42,0.58)$. 


\section{Models}\label{sec:model_all}

\subsection{Fitting procedure}

We fit our gradient tree boosting models with the approach outlined in \S\ref{sec:model:boosting}, minimizing the loss functions described in \S\ref{sec:lossfunctions}. 

We use the Poisson and dGPD loss functions to fit models on the full CNT distribution. We experimented with different high thresholds $u$ but achieved the best prediction when we modelled the full distribution with the dGPD, i.e. $u=0$. \JON{For all observations with a masked count response but an unmasked zero valued burnt area, we artificially set their corresponding counts to zero, and use these observations to fit the CNT models.}

For wildfire sizes, we fit models with the squared loss on log(1+BA$_i$), $i=1,\dots,n,$ \JON{and call the best fitted model from this category `log-Normal'.} We also consider mixture models which require split modelling of the distribution. For these, we first choose a sufficiently high ($95\%$ empirical quantile) threshold at $200$ac. We then use the fire sizes above the threshold to fit a model with the GPD loss, and the log-transformed positive sizes below the threshold with the truncated gamma loss. Lastly, we fit a multi-class classifier to the wildfire size component indicators $\bm y_i= (y_{i,1}, \dots, y_{i,3})$, defined in \S\ref{sec:size:classification}, using the cross-entropy loss from (\ref{eq:cross_entropy}); here, $y_{i,1}=1$ if we observe no fire, $y_{i,2}=1$ if BA$_i$ is a medium fire (between $0$ to $200$ac), and $y_{i,3}=1$ if it is a large fire ($>200$ac). \JON{For the classifier, we also used all the observations with a masked burned area but an unmasked zero valued count, i.e., we artificially set the corresponding burned area to be zero for these observations.} 

Given the covariates at the $i$-th observation $\bm x_i$, we combine the three model components to get the prediction of the cumulative conditional probability for each observation 
\begin{align*}
    \hat{\text{Pr}}(\text{BA}_{i}\leq b \mid \bm x_i) = &\hat{\text{Pr}}({y}_{i,1}=1 \mid \bm x_i) + \hat{\text{Pr}}({y}_{i,2}=1 \mid \bm x_i) \hat{\text{Pr}}(\text{BA}_{i}\leq b \mid \bm x_i, {y}_{i,2}=1) \\ &+ 
\hat{\text{Pr}}({y}_{i,3}=1 \mid \bm x_i) \hat{\text{Pr}}(\text{BA}_{i}\leq b \mid \bm x_i, {y}_{i,3}=1), \quad b \geq 0.
\end{align*}
We also engineered new covariates to improve our predictions. To incorporate more spatial information from our covariates (other than the longitude and latitude coordinates), we took the average value of the covariate across neighbouring grid cells for each month; this smooths the climatic variables across space. We also allow land-use covariates at neighbouring grid cells to help predictions at each grid cell. 

The relationship between CNT and BA is positive, 
\JON{though a high CNT does not imply a high BA}. Nevertheless, it is still natural to consider the other response as a covariate when modelling a given response, or at least to use this information whenever possible. As the test grid cells for BA and CNT are correlated, there are instances where the BA response was masked but the CNT wasn't, and vice versa; for $39\%$ of masked BA observations, their corresponding CNT observations were unmasked. Using the CNT/BA covariate to predict BA/CNT thus raises the question of how best to impute its value if it was masked for a given observation. The default way to handle a missing covariate is to impute its mean across all observations, such as in algorithms from \texttt{xgboost} or \texttt{gbm}, though this will be sub-optimal for predictions on a spatially heterogenous dataset. Instead, we use an imputation method which fits a model for the covariate and then imputes the best estimate from this model. 

When modelling BA, we first fit a gradient boosting model with the dGPD loss on the CNT response, without using BA as a covariate \JON{so as to prevent data leakage}, and then use the estimated parameters to find the estimated mean CNT for each observation using (\ref{eq:dgpd:mean}); we then impute this estimate whenever CNT was masked. When modelling CNT, we first fit a gradient boosting model, without using CNT as a covariate, with the cross-entropy loss on the wildfire size component indicators $\bm y_i$, $i=1,\dots,n$, and then impute the estimates of the probabilities $\text{Pr}(y_{i,1}=1)$, $\text{Pr}(y_{i,2}=1)$ and $\text{Pr}(y_{i,3}=1)$ from the fitted model, whenever BA was masked; when BA wasn't masked, we use the observed indicators $\bm y_i$ as a covariate. 

To assess models using these covariates, a cross-validation scheme should also reflect this imputation procedure; thus, it becomes even more important to appropriately model the inter-variable dependence between CNT and BA of the masking processes, i.e., the parameter $\beta$ in our spatiotemporal cross-validation scheme described in \S\ref{sec:methods:cv}. 

Our models have hyperparameters that must be tuned by cross-validation. They include the regularization parameters $\lambda$ and $\eta$ in \JON{(\ref{eq:impurity})}, the proportion $s$ of covariates subsampled at each iteration, the maximum number of leaves for each tree $L$, and the number of trees $T$ (see \S\ref{sec:model:boosting}). Other hyperparameters from the loss functions include $k$, $\xi$, and $\alpha$, which govern the shape and tails of the fitted conditional distribution, and weights $w_i$ $(i=1,\dots,n)$ which govern the importance of each observation in the cross entropy loss. Some of our models assume common shape parameters $\xi$ and $\alpha$ governing the tails of wildfire sizes and counts across the whole sample space; our preliminary analysis suggests that this assumption is reasonable in space, as the frequentist estimates of the shape parameters from pooled data are relatively homogeneous across the wildfire coordination regions.

We use the Bayesian optimization procedure outlined in \citet{Snoek.2012} to choose the parameters (excluding the number of trees $M$) minimizing the average evaluation metric, calculated on the five cross-validation folds generated in \S\ref{sec:methods:cv}. This procedure 
treats the objective function as random and first places a Gaussian process prior on it. After gathering function evaluations, the prior is updated to form the posterior distribution over the objective function, which is then used to construct an infill sampling criterion. 
For the mixture model, we implement separate Bayesian optimization procedures for each of the three model components.

\begin{figure}[t]
    \centering
\includegraphics[width=.7\textwidth]{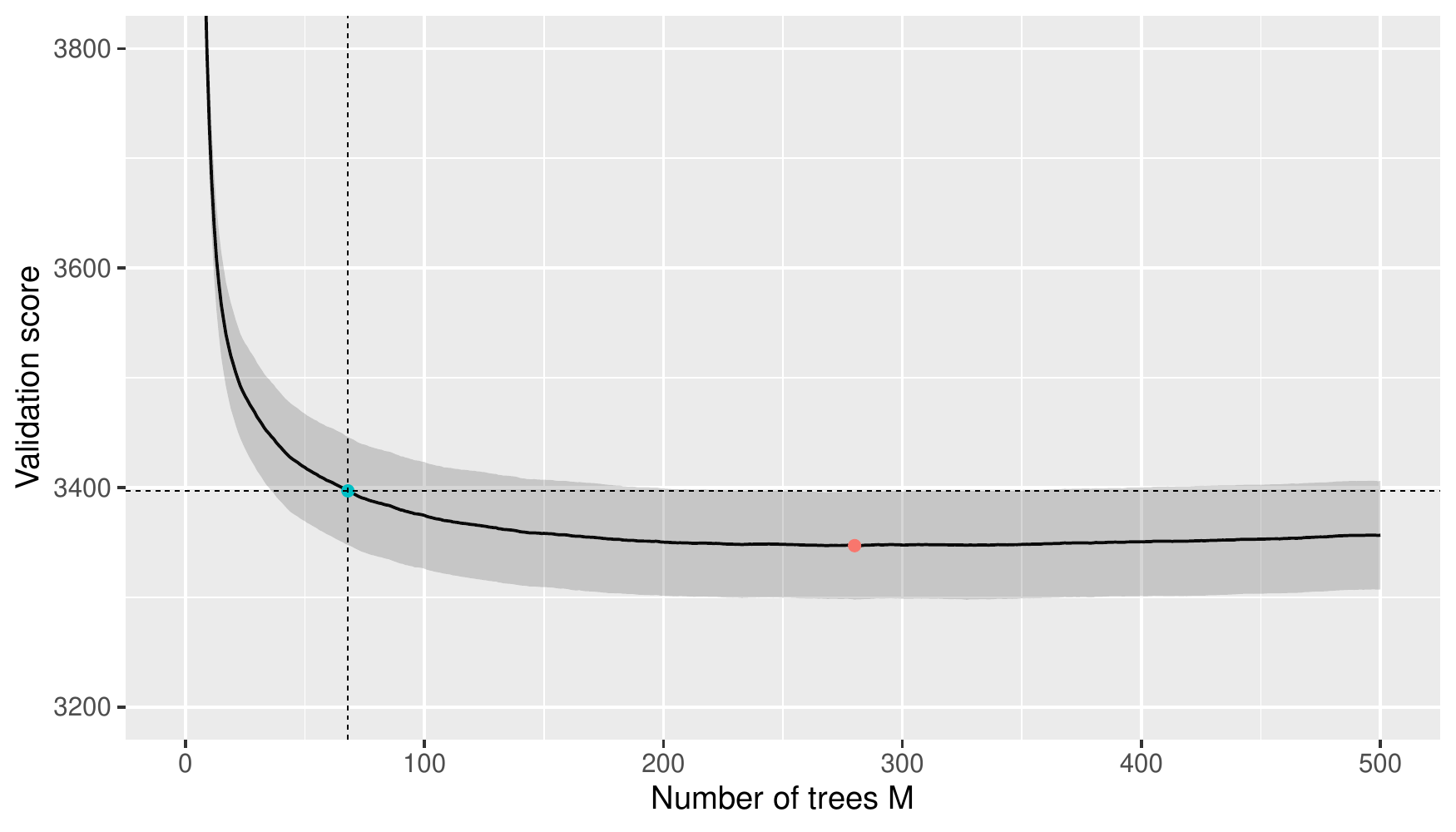}
  \caption{The average rescaled evaluation score across all five folds of our spatiotemporal cross-validation scheme for a mixture model, as a function of the number of trees $M$. The shaded region shows the pointwise one-standard-error bound. The red point shows the minimum average validation score and the blue point shows the $M$ chosen by the one-standard-error rule. }
  \label{fig:CV_M}
\end{figure}

Given the other parameters, we choose $M$ with the one-standard-error rule \citep{hastie.2009}; i.e., we select the largest $M$ within one standard error of the parameter that achieves the minimum in terms of the evaluation metric. Figure \ref{fig:CV_M} shows the evaluation metric on the validation folds as a function of $M$ for the wildfire size classifier in a mixture model. 

We fitted our models by combining our own \texttt{R} routines with the \texttt{xgboost} package. We implemented the described Bayesian optimisation procedure with the \texttt{rBayesianOptimization} package. 




\subsection{Results}\label{sec:results_boosting}

The benchmark models are described in \citet{data.challenge}. That for CNT corresponds to a generalized linear model \citep[GLM,][]{Nelder-Wedderburn.1972} with a Poisson response distribution and log link linear predictor using all the original covariates. The benchmark for BA first fits a generalized linear model with Gaussian response and log-link using all of the original covariates. The probability predictions, $\text{Pr}(\text{BA}_i \leq u_{\text{BA}})$, are obtained by combining the log-Gaussian BA model with the estimated probability that CNT$_i$=0, obtained from the benchmark Poisson model for CNT. 

We relied on our cross-validation scheme devised in \S\ref{sec:methods:cv}, using the evaluation metrics used from the competition outlined in \S\ref{sec:data_boosting}, to choose which model predictions to use for the competition. After the competition, we had access to the truth and could calculate how the predictions of every model would have performed on the test set. 


Table \ref{table:clic_simulation} shows that incorporating the engineered CNT and BA covariates improves the scores of all models by up to $7\%$. According to our cross-validation scheme, the best model for wildfire sizes is the mixture model, and for the counts it is the dGPD model. The best mixture model and dGPD models from the Bayesian optimization procedure have $\alpha=52$ and $\xi=0.8$. This implies a fat tail for the size distribution, but a thinner (Gumbel-like) tail for the wildfire count distribution, though the parameter $\alpha$ in the dGPD loss provides additional flexibility to the model that gives slightly better predictions than the Poisson model. All our gradient boosting models outperform the benchmark by around $10$--$50\%$.

Our cross-validation scheme tends to perform better than the $5$-fold cross-validation scheme as a proxy for the true test set performance; the scores from the $5$-fold cross-validation scheme are generally too optimistic compared to the true test error, especially for the wildfire size models. This optimism is especially pronounced when evaluating models that use the engineered CNT and BA covariates. Our scheme is better able to capture the inter-variable dependence between the CNT and BA masking processes, giving a better reflection of how the models that incorporate the engineered covariates would perform when predicting responses on the test set.

\begin{table}[t]
\centering
\begin{tabular}{lrrrr}
 & Model & \JON{Data-driven CV} & $5$-fold CV & Truth \\ 
  \hline
 \multirow{2}{*}{Counts} & Benchmark$^*$ & 5172 & 5235 &   5565\\ 
 & Poisson$^*$ & 3413 & 3213 & 3302 \\ 
& Poisson & 3283 &  3102 &   3131 \\  
& dGPD$^*$ & 3304  & 3092 & 3194  \\ 
& dGPD & $\mathbf{3215}$  & 2896 & 3068  \\ 
  \hline
     \multirow{2}{*}{Sizes} & Benchmark$^*$ & 3923  & 3834  & 4244 \\
             & Log-Normal$^*$ & 3553 & 3450 & 3771  \\ 
        & Log-Normal & 3473 & 3202 & 3501  \\  
    & Mixture$^*$ & ${3480}$ & 3403 & 3582  \\  
& Mixture & $\mathbf{3364}$ & 3089 & 3446  \\ 
     \hline
\end{tabular}
\caption{The averaged rescaled evaluation score for all models, according to the data-driven cross-validation (CV) scheme outlined in \S\ref{sec:methods:cv}, the $5$-fold CV scheme with random partitioning, and the true score. The bold figures highlight the best model chosen by our cross-validation approach. The asterisks indicate models not using the CNT/BA covariate. }
\label{table:clic_simulation}
\end{table}

\begin{figure}[t]
\centering
  \begin{subfigure}[b]{.45\linewidth}
    \centering
\includegraphics[width=.99\textwidth]{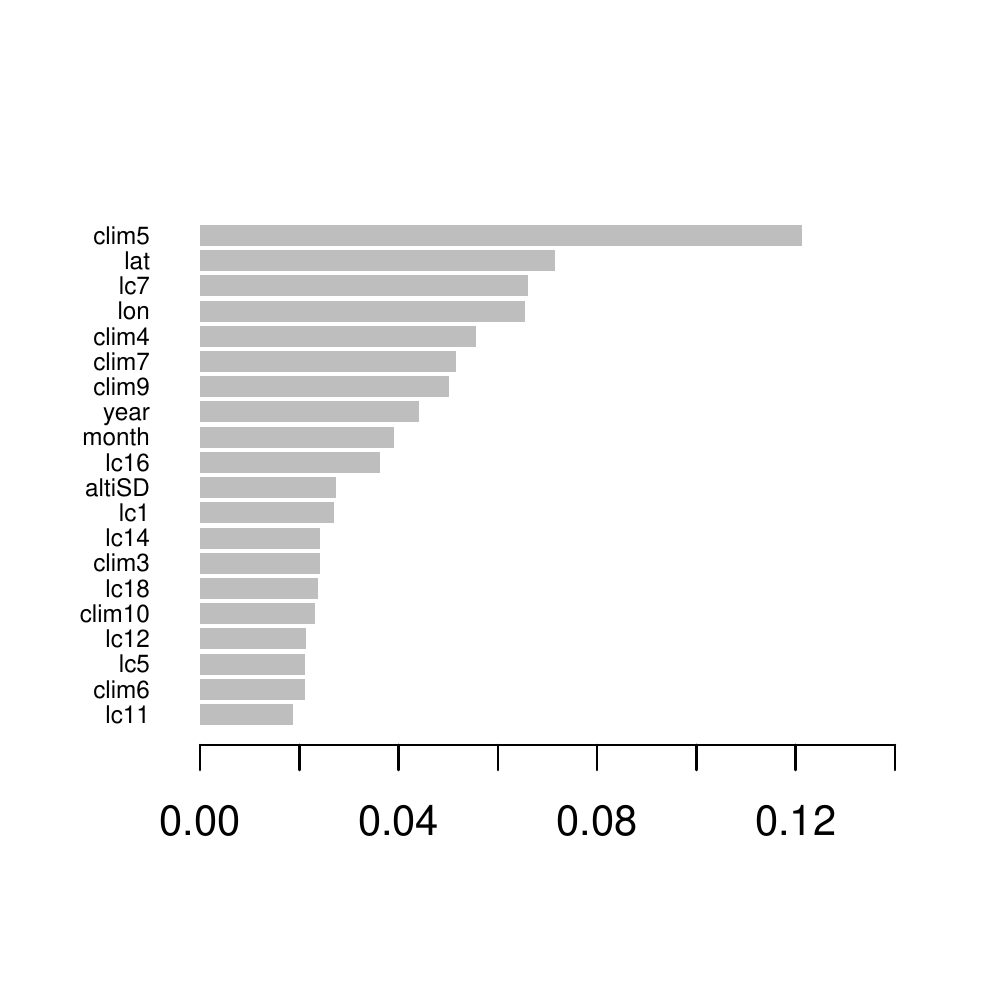}
  \end{subfigure}
  \begin{subfigure}[b]{.45\linewidth}
    \centering
    \includegraphics[width=.99\textwidth]{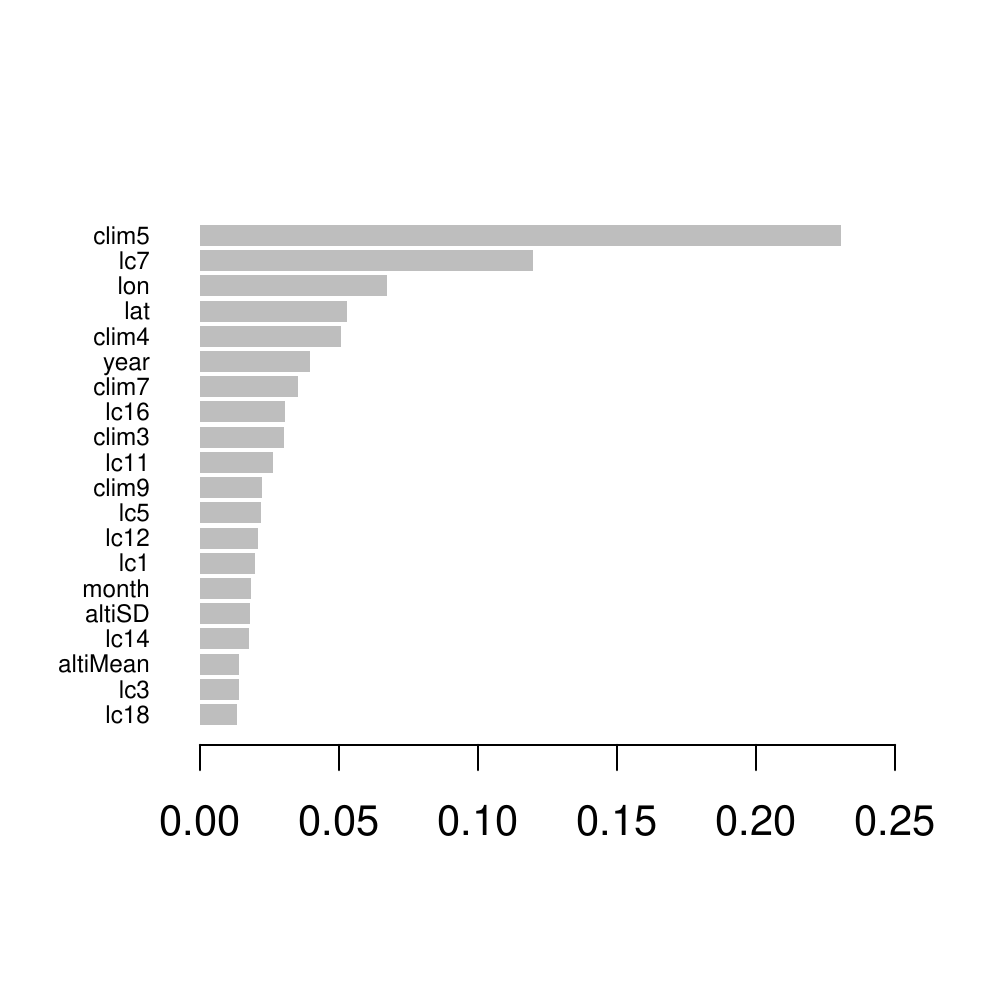}
  \end{subfigure} 
  \caption{The coverage (left) and gain (right) metrics for the top 20 covariates in the wildfire size classifier submodel of the best mixture model for BA (without using the engineered covariates). More information about the covariates is given in \citet{data.challenge}.}
  \label{fig:variable_importance}
\end{figure}

Figure \ref{fig:variable_importance} shows that the gain and coverage metrics introduced in \S\ref{sec:model:boosting} give similar orderings for the importance of covariates when predicting the probability of being in a given wildfire size component with the best mixture model. As hypothesized in \S\ref{sec:data_boosting}, clim5, lc7 and the spatial covariates of longitude (lon) and latitude (lat) are among the five most important variables for both metrics. The other variables (e.g., clim4, clim7, clim9, year, lc16, etc.) are relevant, but each is less than half as important as clim5, the most important covariate.


To evaluate the marginal effect of clim5 on the CNT response in the best dGPD model, we transformed the partial dependence estimate (\ref{eq:pdp}) by (\ref{eq:dgpd:mean}) to get the predicted mean count $\hat{m}_i$, and evaluated and plotted it with clim5 in the set of interest $\mathcal{S}$ in (\ref{eq:pdp}). As it is computationally infeasible to evaluate all data points $\bm x_{i\mathcal{C}}$ in our setting with over $500,000$ observations, we subsampled $10,000$ observations to obtain our Monte Carlo estimates.

\begin{figure}[t]
\centering
  \begin{subfigure}[b]{.45\linewidth}
    \centering
\includegraphics[width=.99\textwidth]{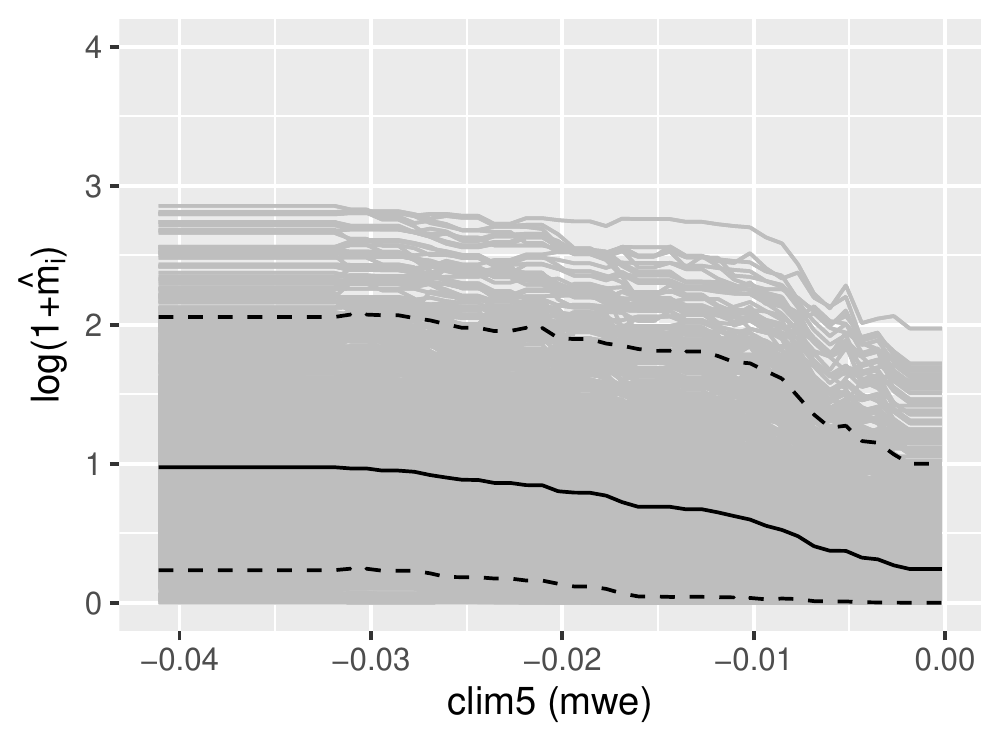}
  \end{subfigure}
  \begin{subfigure}[b]{.45\linewidth}
    \centering
    \includegraphics[width=.99\textwidth]{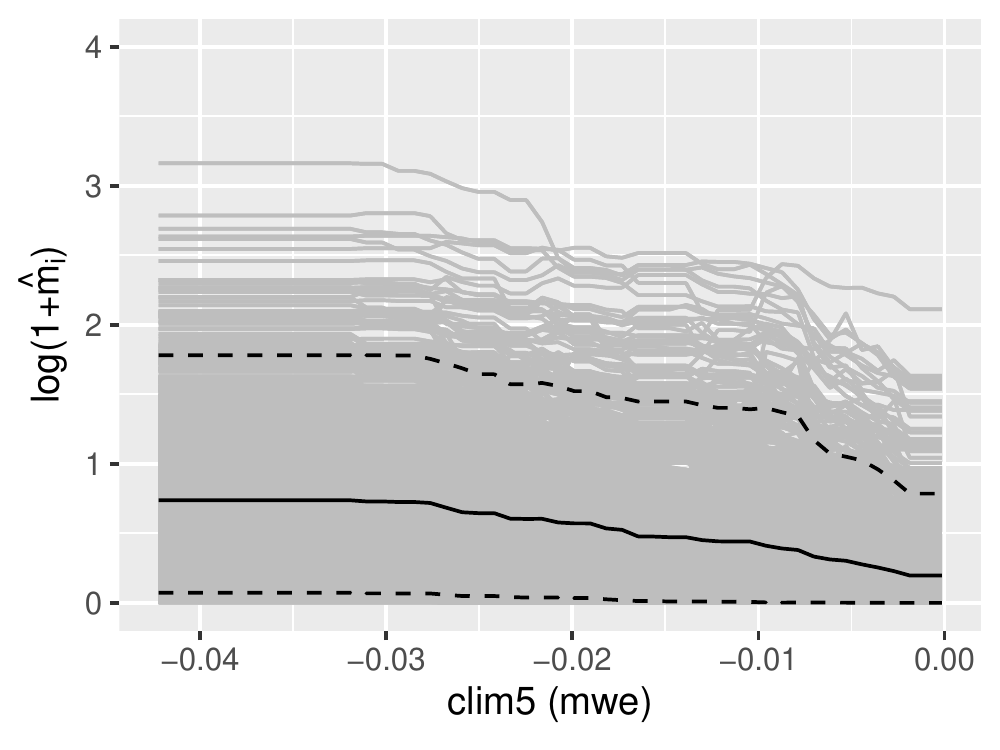}
  \end{subfigure} 
  \caption{The partial dependence plot with clim5 in the set of interest $\mathcal{S}$, transformed so the $y$-axis shows the predicted mean CNT from (\ref{eq:dgpd:mean}), for the observations in the Rocky Mountain Area (left) and Great Basin (right) regions. The empirical $95\%$ and $5\%$ pointwise quantiles from the Monte Carlo estimates are shown by the dotted lines. }   
  \label{fig:pdp}
\end{figure}

\begin{figure}[t]
\centering
  \begin{subfigure}[b]{.8\linewidth}
    \centering
\includegraphics[width=.99\textwidth]{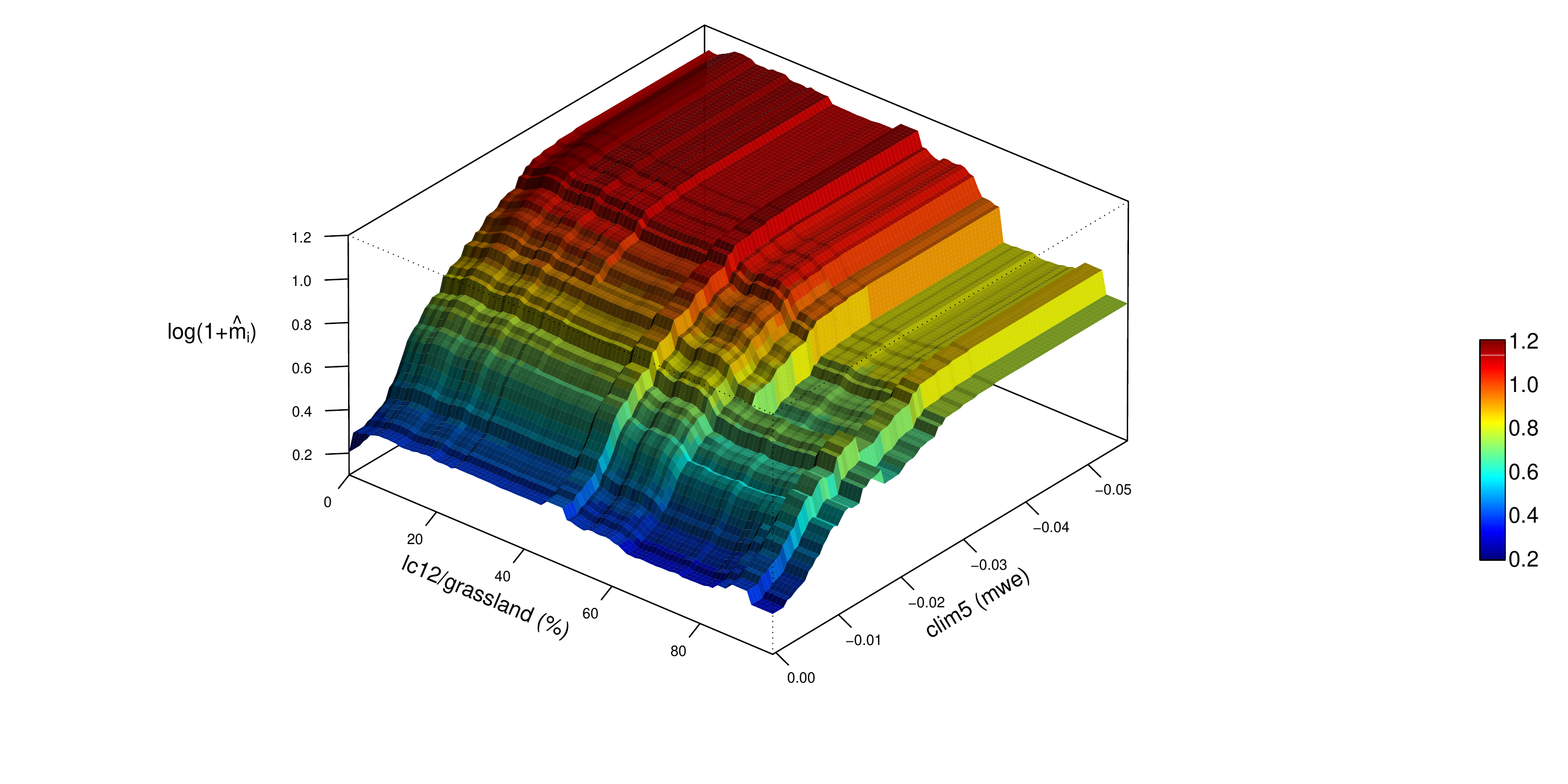}
  \end{subfigure}
  \caption{The three-dimensional partial dependence function with clim5 and lc12 in the set of interest $\mathcal{S}$, transformed as in Figure \ref{fig:pdp} and plotted for all observations. }   
  \label{fig:pdp:interaction}
\end{figure}

Figure \ref{fig:pdp} shows that the marginal effect of clim5 on $\hat{m}_i$ tends to be negative, especially above $-0.03$mwe. Figure \ref{fig:pdp:interaction} displays the joint marginal effects of clim5 and land cover covariate 12 (lc12; grassland, in $\%$) on the predicted mean count, i.e., with clim5 and lc12 in the set of interest $\mathcal{S}$ in (\ref{eq:pdp}). The figure hints at interaction between the two covariates; increasing lc12 tends to decrease the response CNT slightly if clim5 is low, but not when clim5 is high. Although the partial dependence plot is useful for showing the overall marginal trend of a covariate on the response across all observations considered, it is important to be honest about the uncertainty associated with the Monte Carlo estimate in (\ref{eq:pdp}). The estimates in Figures \ref{fig:pdp} and \ref{fig:pdp:interaction} are associated with high uncertainty throughout (not shown for the latter); our dataset is very heterogeneous and it is not possible to quantify the marginal effects of covariates with less uncertainty. 


Our chosen models perform competitively when compared to the other teams' submission in the data challenge \citep{data.challenge}, placing second out of 28 teams in the final ranking; the other top three teams used other popular prediction techniques such as random forests, hierarchical Bayesian modelling and ANN models with adapted loss functions \citep{Opitz.2022}. 





\section{Discussion}\label{sec:conclusion_boosting}

We have implemented novel gradient boosting models for wildfire activity that are trained with loss functions motivated by extreme-value theory. Compared to models trained on the Poisson loss, our chosen model for wildfire counts has an additional parameter $\alpha$ that governs the tail of the count distribution, which, after tuning by cross-validation, enables the model to give better predictions. Our chosen model for burned areas has specific components for extreme fire sizes. According to the given score criteria which put more weight on large fire sizes, this model improves on the models that do not discriminate between extreme and non-extreme fire sizes. 

As the use of other data sources, e.g., covariates not included in the provided dataset, was strictly prohibited for the competition, we were not able to leverage other spatial information, such as the Geographic Area Coordination Center of a grid cell. Each center follows its own governing jurisdictions that could affect its fire mitigation and suppression strategies. Including this information, either by incorporating additional covariates, or by having a separate gradient boosting model for each coordination center, would improve predictions.

Our mixture model has the threshold fixed at $u=200$ac. However, $u$ could have also been allowed to \JON{be} an additional parameter chosen by cross-validation, or as a spatio-temporally varying threshold $u_{d,m}$, $d\in\{1,\dots,D\}$, $m\in\{1,\dots,M\}$, \JON{that could be either independent \citep{Opitz.al.2018} or dependent \citep{velthoen.2021} on covariates. However, implementing these approaches would significantly raise the computational cost here, especially if the model for the threshold also has hyperparameters that need to be tuned}, as the optimization of all of our model components would need to be done jointly because they all rely on the threshold. \JON{Future work could investigate whether these potential approaches that increase complexity do indeed improve wildfire prediction.}

Our best chosen dGPD model from the Bayesian optimization procedure has $\alpha=52$, which implies a thin tail for the CNT distribution that is not too different from the Poisson distribution. This explains the slight improvement of the model using the dGPD compared to the Poisson loss. Had the tail of the CNT distribution been fatter, as in the case of another data application (e.g., insurance claim counts), we would have also noticed a larger improvement in predictions by the dGPD model.

We have implemented a spatial cross-validation scheme for our context which partly fixes the optimism when using traditional $k$-fold cross-validation to evaluate complex models with engineered covariates over a spatially heterogeneous but dependent dataset. 
One should always have the real-world prediction scenario in mind when choosing a cross-validation scheme, and we appealed to tools from spatial statistics to aid the validation of model predictions on our test data.


Apart from cross-validation approaches, model comparison using other predictive scores, e.g., Continuous Ranked Probability Scores \citep[CRPS,][]{Matheson-Winkler.1976} or tail-weighted CRPS \citep[][]{Gneiting-Ranjan.2011}, could be used to compare simulated predictive distributions of burned areas or counts. These scores, along with graphical summaries for validating models, are an important future topic of research, and have already been explored in the extreme wildfire prediction context by \citet{Joseph2019}, \citet{Pimont2020} and \citet{Koh.2021}. Due to the time constraints of the competition however, this is out of the scope of this paper. 

Our gradient boosting models have hyperparameters from the loss functions that govern the tails of the predictive distributions: $\xi$, and $\alpha$, which, once chosen by cross-validation, are fixed in the model. To allow more flexible modelling of the distributional tails, one could incorporate them as an additional boosting estimate in \S\ref{sec:model:boosting} which would allow these parameters to depend on the covariates. The boosting estimate $\hat{\theta}_i$, gradient $g_i$ and hessian $h_i$ in (\ref{eq:objective}) would then be two-dimensional vectors. The described approach is similar to the recent work by \citet{velthoen.2021}. 
\JON{Another avenue for future work is to assess how the the tail indices affect the cross-validation score when the other hyperparameters are fixed.}



Apart from better predictions, our models improve decision support in wildfire management. The partial dependence plots in Figures \ref{fig:pdp} and \ref{fig:pdp:interaction} allow marginal and interaction effects of covariates to be assessed, though one should be aware of the large uncertainty associated with these estimates. Importance metrics like the gain and coverage in Figure \ref{fig:variable_importance} could be used for covariate selection and could prompt national wildfire predictive services to rethink designs of fire danger warning systems (e.g., indices) across the contiguous US.

\begin{acknowledgements}
The author is grateful to the two anonymous referees and the Guest Editor 
for their constructive comments that improved the quality of this paper. Further thanks go to Anthony C. Davison for helpful comments and discussions, Thomas Opitz for organising the EVA 2021 data challenge, and the Swiss National Science Foundation for financial support. 
\end{acknowledgements}

%
\section*{Conflict of interest}

The author declares that he has no conflict of interest.

\section*{Data availability statement}

The data that support the findings of this study are from the EVA 2021 data challenge \citep{Opitz.2022}, and are available on request from the the organiser.

\bibliographystyle{agsm_jon.bst}      
\bibliography{reference.bib}   


\newpage

\section{Supplement}\label{sec:appendix}

\subsection{Terms in the gradients and hessians}

\subsubsection{For the truncated gamma loss}

The lower incomplete gamma function is $\gamma(k,s) = \int_0^{s} t^{k-1} \exp(-t) \text{d}t$, $s>0$. Set $s=ku/\exp(\hat{\theta}_i)$. By applying the fundamental theorem of calculus and the chain rule, taking derivatives with respect to $\hat{\theta}_i$ gives \JON{
\begin{align*}
\gamma'\{k, k u/\exp(\hat{\theta}_i)\} = 
& \exp(\hat{\theta}_i) \left\{ -\left(\dfrac{ku}{\exp(\hat{\theta}_i)}\right)^{k-1}\exp\left(-\dfrac{ku}{\exp(\hat{\theta}_i)}\right) \dfrac{ku}{\exp(\hat{\theta}_i)^2} \right\}, \\ &u>0, \hat{\theta}_i \in \mathbb{R},
\end{align*}
and 
\begin{align*}
\gamma''\{k, k u/\exp(\hat{\theta}_i)\} =  & \Bigg[ \Bigg\{ \left(\dfrac{ku}{\exp(\hat{\theta}_i)^2}\right)^2 (k-1) \left(\dfrac{ku}{\exp(\hat{\theta}_i)}\right)^{k-2} \\
&+2 \left(\dfrac{ku}{\exp(\hat{\theta}_i)}\right)^{k-1} \dfrac{ku}{\exp(\hat{\theta}_i)^3}  \Bigg\} \exp\left(-\dfrac{ku}{\exp(\hat{\theta}_i)}\right) \\ & -\left(\dfrac{ku}{\exp(\hat{\theta}_i)}\right)^{k-1} \left(\dfrac{ku}{\exp(\hat{\theta}_i)^2}\right)^2  \exp\left({\dfrac{-ku}{\exp(\hat{\theta}_i)}}\right) \Bigg] \exp(\hat{\theta}_i)^2 \\& +  \gamma'\{k, k u/\exp(\hat{\theta}_i)\}.
\end{align*}
}

\subsubsection{For the GPD loss}

Let $\kappa \in (0,1)$. We first reparameterize the GPD probability density function as
\begin{align}\label{eq:reparam:gpd}
f\{y_i,\exp(\hat{\theta}_i), \xi\} &=   \JON{\dfrac{ \{(1-\kappa)^{-\xi}-1\} }{ \xi \exp(\hat{\theta}_i)}} \Bigg\{1+ \dfrac{y_i \{(1-\kappa)^{-\xi}-1\} }{  \exp(\hat{\theta}_i)} \Bigg\}^{-(\xi+1)/\xi}, \nonumber \\ & \xi>0, \hat{\theta}_i \in \mathbb{R}.
\end{align}
The functions $f'\{y_i,\exp(\hat{\theta}_i), \xi\}$ and $f''\{y_i,\exp(\hat{\theta}_i), \xi\}$ are obtained by differentiating (\ref{eq:reparam:gpd}) with respect to $\hat{\theta}_i$. Write $A= \{(1-\kappa)^{-\xi}-1\} /  \exp(\hat{\theta}_i)$, and notice that $\partial A/\partial \hat{\theta}_i = A' = -A$. Then \JON{
\begin{align*}
    f'\{y_i,\exp(\hat{\theta}_i), \xi\} = {\dfrac{A^2y_i(\xi+1)}{\xi^2}}\left(1+Ay_i\right)^{-({2\xi+1)/\xi}}-f\{y_i,\exp(\hat{\theta}_i), \xi\},
\end{align*}
and the} expression for $f''\{y_i,\exp(\hat{\theta}_i), \xi\}$ can be easily obtained in a similarly straightforward manner.

\subsection{Priors and SPDE triangulation}


This section details the prior and SPDE triangulation specifications of the spatiotemporal Bernoulli model for the masking processes.

The fixed effect coefficient $\beta_0^\mathrm{CNT}$ and $\beta_0^\mathrm{BA}$ in our model was assigned a flat Gaussian prior with zero mean and precision $0.001$. The prior for the scaling parameter $\beta$ is a zero-centered Gaussian distribution with precision $\omega=1/20$. Lastly, we assigned a log-gamma hyperprior with mean unity and precision $0.0005$ to the variance hyperparameter $\phi$.

The spatial Gaussian random effects $g_m$ and their  conditional distributions must be tractable in our setting with many pixels. We use the Mat\'ern covariance function for random effects, given  as follows for two points $s_1$ and $s_2$:
$$
\mathrm{Cov}\{g_m( s_1), g_m( s_2)\} = \sigma^2 2^{1-\nu} (\kappa||  s_1 -  s_2||)^{\nu} K_{\nu}(\kappa ||  s_1 -  s_2||) /\Gamma(\nu), \quad \sigma, \nu >0,
$$
with Euclidean distance $||\cdot||$, gamma function $\Gamma$, the Bessel function of the second kind $K_{\nu}$, and standard deviation and smoothness parameters $\sigma$ and $\nu$. The {empirical range}, at which the correlation drops to approximately $0.1$, is $r=\sqrt{8\nu}/\kappa$.
Numerically convenient representations by approximating Gauss-Markov random fields (GMRF, characterized by sparse precision, \ie inverse covariance, matrices) are constructed by solving a stochastic partial differential equation driven by Gaussian noise \citep[SPDE,][]{Lindgren.al.2011}, where we fix the smoothness $\nu$ at unity.  

The parameter vector $\bm \zeta$ consists of $r$ and $\sigma$ parameters which were assigned penalized complexity priors \citep{Fuglstad.al.2018}. Such priors penalize the distance of the prior of a model component towards a simpler baseline at a constant rate \citep{Simpson.al.2017}. 

\begin{figure}[t]
\centering
\begin{subfigure}{.7\linewidth}
    \includegraphics[width=.99\textwidth]{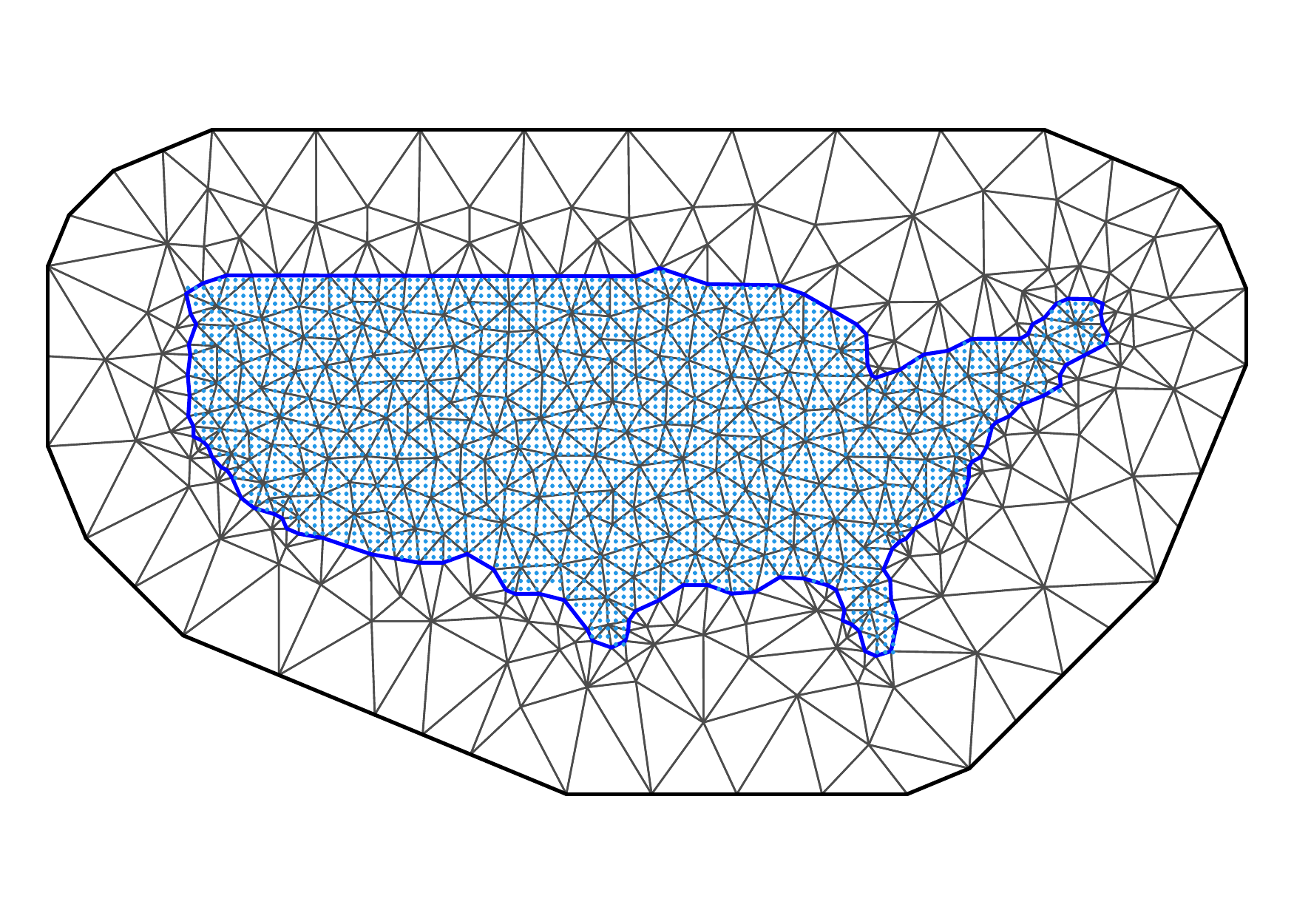}
\end{subfigure}
  \caption{Triangulation mesh of the spatial study region (blue contours) for the SPDE approach. Neumann boundary conditions are set on the exterior (black) boundary to obtain a unique solution. The blue points represent the centroids of the grid cells. The finite element solution defines a Gauss--Markov random vector with one variable in each node.}
    \label{fig:mesh_boosting}
\end{figure}

The discretization points when triangulating the spatial domain are chosen as the nodes of a finite element representation which enables efficient inference for random effects representing spatial variation. Our spatial triangulation mesh in Figure~\ref{fig:mesh_boosting} has $508$ nodes. It is sparser in the extended zone around the study area to ensure that the SPDE boundary conditions have negligible influence on the study area.

For computational reasons, we only use observations of the masking processes from the first ten months, and subsample $30\%$ of the data when fitting our model.

\end{document}